\documentclass[aps,twocolumn,prb,showpacs]{revtex4}

\usepackage[dvips]{color}
\usepackage{graphicx}
\usepackage{epsfig}
\usepackage{dcolumn}
\usepackage{amssymb}
\usepackage[english]{babel}
\usepackage{here}
\usepackage{mathrsfs}
\usepackage{bm}
\usepackage{amsmath}
\usepackage{lscape}

\begin{document}

\title{Photonic-crystal slabs with a triangular lattice of triangular holes
investigated \\ using a guided-mode expansion method}

\author{Lucio Claudio Andreani and Dario Gerace$^*$}

\affiliation{Dipartimento di Fisica ``A. Volta,'' Universit\`a di
Pavia, via Bassi 6, I-27100 Pavia, Italy}
\date{\today}

\begin{abstract}
According to a recent proposal [S. Takayama \textit{et al.}, Appl.
Phys. Lett. \textbf{87}, 061107 (2005)], the triangular lattice of
triangular air holes may allow to achieve a complete photonic band
gap in two-dimensional photonic crystal slabs. In this work we
present a systematic theoretical study of this photonic lattice in
a high-index membrane, and a comparison with the conventional
triangular lattice of circular holes, by means of the guided-mode
expansion method whose detailed formulation is described here.
Photonic mode dispersion below and above the light line, gap maps,
and intrinsic diffraction losses of quasi-guided modes are
calculated for the periodic lattice as well as for line- and
point-defects defined therein. The main results are summarized as
follows: (i) the triangular lattice of triangular holes does
indeed have a complete photonic band gap for the fundamental
guided mode, but the useful region is generally limited by the
presence of second-order waveguide modes; (ii) the lattice may
support the usual photonic band gap for even modes (quasi-TE
polarization) and several band gaps for odd modes (quasi-TM
polarization), which could be tuned in order to achieve
doubly-resonant frequency conversion between an even mode at the
fundamental frequency and an odd mode at the second-harmonic
frequency; (iii) diffraction losses of quasi-guided modes in the
triangular lattices with circular and triangular holes, and in
line-defect waveguides or point-defect cavities based on these
geometries, are comparable. The results point to the interest of
the triangular lattice of triangular holes for nonlinear optics,
and show the usefulness of the guided-mode expansion method for
calculating photonic band dispersion and diffraction losses,
especially for higher-lying photonic modes.
\end{abstract}



\maketitle

\section{Introduction}

Photonic crystals (PhC) embedded in planar waveguides, also called
PhC slabs, are at the heart of current research on photonic
crystals\cite{joanno,sakoda,johnson,inoue,busch,lourtioz} because
of the possibility of confining light in all spatial directions
combined with the advantage of a lithographic definition of the
pattern. The propagation of light in these systems can be
controlled by a two-dimensional (2D) photonic lattice in the
waveguide plane, and by total internal reflection in the
perpendicular direction. Waveguides with strong refractive index
contrast (like the suspended membrane, or air bridge) support
truly guided modes lying below the light line dispersion in the
cladding
materials.\cite{russell95,krauss96,johnson99,chutinan00,johnson00,qiu02,notomi02,whittaker02}
However, most of the photonic modes (or all of them, in the case
of waveguides with weak refractive index contrast) lie above the
light line and are only quasi-guided, as they are subject to
intrinsic losses due to out-of-plane
diffraction.\cite{lalanne01,palamaru01,ochiai01-fdtd,ochiai01-pert,lalanne02,
hadley02,fan02,tikhodeev02,morand03,ferrini03,cryan05}

The physical properties of PhC slabs can be significantly
different from those of the corresponding 2D system, for several
reasons. First, the 2D photonic modes are subject to confinement
in the vertical waveguide and the resulting blue shift is strongly
polarization-dependent. Second, the presence of second- and
higher-order waveguide modes can produce a complicated pattern of
photonic bands, especially for higher-lying states. Third,
diffraction losses of quasi-guided modes are an inherent feature
of PhC slabs which is absent in the ideal 2D case. For all of
these reasons, some basic and well-known properties of 2D photonic
crystals cannot be easily translated to PhC slabs. For example, it
is well known that the triangular lattice of circular air holes in
2D supports a complete photonic band gap for all propagation
directions and light polarizations at sufficiently large air
fractions.\cite{joanno,meade92,villeneuve92,gerard94}
Nevertheless, the same lattice realized in a high-index membrane
does not possess a complete band gap, as the odd modes with
respect to a horizontal mirror plane (often called quasi-TM modes)
are subject to a strong and non-uniform blue shift which
eliminates the gap.\cite{johnson99,andreani02_ieee} Indeed, most
applications of PhC slabs employ the even modes (often called
quasi-TE), which do possess a band gap for all propagation
directions.

It was recently suggested\cite{takayama05} that the triangular
lattice of triangular air holes in a high-index membrane gives
rise to a complete photonic gap for both even and odd modes. The
physical mechanism is the reduction of symmetry of the basis in
the unit cell, as compared to the hexagonal symmetry of the 2D
lattice, giving rise to a splitting between the first and the
second odd bands at the K point of the Brillouin zone: when
realized in a high-index membrane, this gap overlaps the usual gap
between the first and the second even bands, giving rise to a
complete gap for all directions and polarizations. The
experimental results reported in Ref.~\onlinecite{takayama05}
support the existence of a complete photonic gap.

A main purpose of this paper is to perform a systematic study of
the triangular lattice of triangular holes, as compared to the
triangular lattice of circular holes, both being realized in a
high-index dielectric membrane. We calculate the photonic band
dispersion, gap maps, and intrinsic losses of quasi-guided modes
for the 2D lattice. We also treat line-defect waveguides obtained
by removing a full row of holes, and point cavities consisting of
three missing holes. In addition to a determination of photonic
gaps for even and odd modes as a function of membrane thickness
and air fraction, we compare diffraction losses for the lattices
with conventional (circular) and reduced symmetry (triangular)
holes: this comparison is important in order to assess the
possible usefulness of the reduced-symmetry lattice. Among the
results, we find interesting prospects of the triangular lattice
of triangular holes for nonlinear optics, as it may allow to
achieve doubly-resonant second-harmonic generation (SHG) with an
even (quasi-TE) fundamental wave and an odd (quasi-TM) harmonic
wave when line-defect waveguides or photonic cavities are
introduced.

The calculations reported in this paper are performed with an
approach which we name Guided-Mode Expansion (GME) method. Maxwell
equations are treated by expanding the magnetic field into the
basis of guided modes of an effective homogeneous waveguide, and
by solving the resulting eigenvalue equation numerically.
Intrinsic losses of quasi-guided modes are obtained by calculating
the coupling to leaky modes of the effective waveguide within
perturbation theory (i.e., the photonic analog of Fermi Golden
Rule for quantum-mechanical problems). The GME method, although
being an approximate one (since the basis of guided modes is not a
complete basis set), has been applied to a variety of photonic
lattices and has proven to be useful especially for obtaining
quasi-guided modes and their diffraction
losses.\cite{andreani02_ieee,andreani02_pss,andreani03_apl,gerace04_pre,
gerace04_pss,andreani04_pn,gerace05_oqe} It has also been
successfully employed for the interpretation of optical
experiments on PhC
slabs.\cite{patrini02,galli02_epjb,malvezzi03,galli04,galli05_jsac,
galli05_prb,gerace05_apl} Another purpose of this paper is to
provide a detailed description of the GME method, together with
convergence tests and exemplifying applications.

The rest of this work is organized as follows. In Section II we
outline the GME method and discuss a few convergence tests.
Section III contains the results for the 2D triangular lattice of
triangular holes in a high-index membrane, namely photonic
dispersion, gap maps and intrinsic losses. In Sec. IV we present a
few results for line-defect waveguides and point cavities in the
triangular lattice of triangular holes. Section V contains a
discussion of the results and of prospective applications of the
investigated lattice. Technical details, which are needed by the
reader in order to implement the GME method, are given in the
Appendices.

\section{Guided-mode expansion method}

As PhC slabs are intermediate between 2D photonic crystals and
dielectric slab waveguides, it is reasonable to describe photonic
modes in these systems starting from slab waveguide modes and
introducing the effect of a dielectric modulation in the core and
cladding layers. This is the central idea of the guided-mode
expansion method, in which PhC slab modes are expanded in the
basis of guided modes of an effective homogeneous waveguide and
coupling to radiative modes is taken into account by perturbation
theory. In this Section we describe the formalism for calculating
mode dispersion and intrinsic losses, for the general case of an
asymmetric PhC slab, and discuss convergence of the method.

\subsection{Formalism for photonic dispersion}
\label{sec:dispersion}

\begin{figure}[t]
\begin{center}
\vspace*{1cm}
\includegraphics[width=0.6\textwidth]{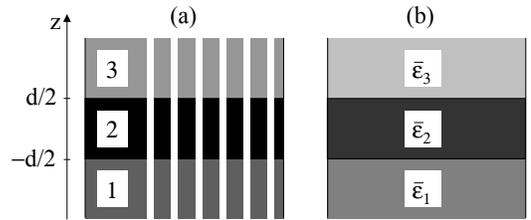}
\vspace*{-6cm} \caption{Schematic picture of the vertical
waveguide structure for (a) the photonic crystal slab and (b) the
effective waveguide. The lower and upper claddings (layers 1 and
3) are taken to be semi-infinite. Photonic patterning in the plane
$xy$ must have the same 2D Bravais lattice for the three layers in
(a), but it can have different bases in the unit cell. The change
in grey scales in going from (a) to (b) suggests that the
effective waveguide is characterized by an average dielectric
constant in each layer.}\label{fig:scheme}
\end{center}
\end{figure}

The system we are considering is shown in Fig.~\ref{fig:scheme}a.
It consists of a PhC slab made of three layers 1,2,3, each of
which is homogeneous in the vertical ($z$) direction and is
patterned with a 2D photonic lattice in the $xy$ plane. The core
layer has a thickness $d$, while the lower and upper claddings are
taken to be semi-infinite.\cite{note:semi-infinite} Patterning of
each layer is characterized by the same 2D Bravais lattice, while
the basis in the unit cell can be different - a typical situation
is that the core layer 2 is patterned, while layers 1 and 3 are
not (i.e., they can be described by the same Bravais lattice as
layer 2, but with a vanishing air fraction). Writing
$\textbf{r}=(\mbox{\boldmath$\rho$},z)$, the dielectric constant
$\epsilon(\mbox{\boldmath$\rho$},z)$ is piecewise constant in the
$z$ direction and takes the form
$\epsilon_j(\mbox{\boldmath$\rho$}),\,j=1,2,3$ in each layer $j$.
We assume a magnetic permeability $\mu=1$.

As is well known, Maxwell equations for the electric and magnetic
fields {\bf E}, {\bf H} with harmonic time dependence can be
transformed into the second-order equation for the magnetic field
\begin{small}
\begin{equation}
 \mbox{\boldmath$\nabla$} \times \left[ \frac{1}{\epsilon({\bf r})}
 \mbox{\boldmath$\nabla$} \times {\bf H} \right] = \frac{\omega^2}{c^2} {\bf H},
 \label{eq:basic}
\end{equation}
\end{small}
with the condition $\mbox{\boldmath$\nabla$}\cdot\textbf{H}=0$. By
expanding the magnetic field in an orthonormal set of basis states
as
\begin{small}
\begin{equation}
{\bf H}({\bf r}) = \sum_{\mu}c_{\mu}{\bf H}_{\mu}({\bf r}),
\label{eq:expansion}
\end{equation}
\end{small}
the orthonormality condition being expressed by
\begin{small}
\begin{equation}
\int {\bf H}_{\mu}^*(\textbf{r})\cdot{\bf
H}_{\nu}(\textbf{r})\,\mathrm{d}\textbf{r}=\delta_{\mu\nu},\label{eq:orthonorm}
\end{equation}
\end{small}
Eq.~(\ref{eq:basic}) becomes equivalent to a linear eigenvalue
problem
\begin{small}
\begin{equation}
\sum_{\nu}{\cal{H}}_{\mu\nu}c_{\nu}=\frac{\omega^2}{c^2}c_{\mu},
\label{eq:linear}
\end{equation}
\end{small}
where the matrix ${\cal{H}}_{\mu\nu}$ is given by
\begin{small}
\begin{equation}
 {\cal{H}}_{\mu\nu}=\int
 \frac{1}{\epsilon({\bf r})}
 (\mbox{\boldmath$\nabla$}\times{\bf H}_{\mu}^*({\bf r})) \cdot
 (\mbox{\boldmath$\nabla$}\times{\bf H}_{\nu}({\bf r})) \, \mathrm{d}{\bf r}.
 \label{eq:matrix-elements}
\end{equation}
\end{small}
This formulation of the electromagnetic problem bears strong
analogies to the quantum-mechanical treatment of electrons, with the
hermitian matrix ${\cal{H}}_{\mu\nu}$ playing the role of a quantum
Hamiltonian. Once the magnetic field of a photonic mode is known,
the electric field can be obtained from
\begin{small}
\begin{equation}
{\bf
E}(\textbf{r})=\frac{ic}{\omega\epsilon(\textbf{r})}\mbox{\boldmath$\nabla$}
\times {\bf H}(\textbf{r}) \label{eq:efield}
\end{equation}
\end{small}
and the electric field eigenmodes are orthonormal according to
\begin{small}
\begin{equation}
\int \epsilon(\textbf{r}) {\bf E}_{\mu}^*(\textbf{r})\cdot{\bf
E}_{\nu}(\textbf{r})\,\mathrm{d}\textbf{r}=\delta_{\mu\nu},
\label{eq:orthonorm-efield}
\end{equation}
\end{small}
as is well known\cite{carniglia71} and easily verified.

In order to define an appropriate basis set ${\bf H}_{\mu}({\bf
r})$ for the expansion (\ref{eq:expansion}), we introduce an
effective slab waveguide made of three homogeneous layers with
dielectric constants $\bar{\epsilon}_1$, $\bar{\epsilon}_2$,
$\bar{\epsilon}_3$, as illustrated in Fig.~\ref{fig:scheme}b. We
take $\bar{\epsilon}_j$ as the spatial average of
$\epsilon_j(\mbox{\boldmath$\rho$})$ in each layer:
\begin{small}
\begin{equation}
\bar{\epsilon}_j=\frac{1}{A}\int_\mathrm{cell}
\epsilon_j(\mbox{\boldmath$\rho$})\,\mathrm{d}{\mbox{\boldmath$\rho$}},
\label{eq:average}
\end{equation}
\end{small}
where the integral extends over a unit cell of area $A$. This
choice for the average dielectric constants is convenient, but by
no means unique - numerical tests and other issues related to the
choice of $\bar{\epsilon}_j$ are discussed in
Subsec.~\ref{sec:convergence}. In any case, we assume that the
average dielectric constants fulfill the inequalities
\begin{small}
\begin{equation}
\bar{\epsilon}_2 > \bar{\epsilon}_1,\bar{\epsilon}_3 \, ,
\label{eq:inequality}
\end{equation}
\end{small}
in order for the effective slab to support a set of guided modes.
Their explicit form is well known\cite{yariv_book} and it will be
summarized here in order to specify our notation, as well as for the
paper to be self-contained. Let us denote by $\textbf{g}=g\hat{g}$
the 2D wavevector in the $xy$ plane with modulus $g$ and unit vector
$\hat{g}$, by $\hat{\epsilon}_{\textbf{g}}=\hat{z}\times\hat{g}$ a
unit vector perpendicular to both \textbf{g} and $\hat{z}$, and by
$\omega_{g}$ the frequency of a guided mode which satisfies
$cg/n_2<\omega_g<cg/\max(n_1,n_3)$. Moreover we define the following
quantities:
\begin{small}
\begin{eqnarray}
\chi_{1g}&=&\left(g^2-\bar{\epsilon}_1\frac{\omega_{g}^2}{c^2}\right)^{1/2},\\
 q_{g}   &=&\left(\bar{\epsilon}_2\frac{\omega_{g}^2}{c^2}-g^2\right)^{1/2},\\
\chi_{3g}&=&\left(g^2-\bar{\epsilon}_3\frac{\omega_{g}^2}{c^2}\right)^{1/2},
\label{eq:chi-and-q}
\end{eqnarray}
\end{small}
which represent the real (imaginary) parts of the wavevector in
the core (upper/lower cladding), respectively. The guided mode
frequencies are found by solving the following implicit equations
(with the suffix $g$ being understood for simplicity):
\begin{small}
\begin{equation}
q(\chi_{1}+\chi_3)\cos(qd)+(\chi_1\chi_3-q^2)\sin(qd)=0
\label{eq:tepol}
\end{equation}
\end{small}
for transverse electric (TE) polarization, and
\begin{small}
\begin{equation}
\frac{q}{\bar{\epsilon}_2}\left(\frac{\chi_1}{\bar{\epsilon}_1}+
\frac{\chi_3}{\bar{\epsilon}_3}\right)
\cos(qd)+\left(\frac{\chi_1\chi_3}{\bar{\epsilon}_1
\bar{\epsilon}_3}-\frac{q^2}
{\bar{\epsilon}_2^2}\right)\sin(qd)=0 \label{eq:tmpol}
\end{equation}
\end{small}
for transverse magnetic (TM) polarization. The guided modes at a
given wavevector \textbf{g} are labelled by the index
$\alpha=1,2,\ldots$ and the eigenfrequencies are denoted by
$\omega_{g\alpha}$. Explicit forms for the electric and magnetic
fields of the guided modes are given in Appendix
\ref{app:guided-modes}, as they are needed for the calculation of
the matrix elements (\ref{eq:matrix-elements}).

The guided modes of the effective waveguide in
Fig.~\ref{fig:scheme}b depend on a wavevector \textbf{g} which can
take any value in the 2D plane. However, photonic modes in the PhC
slab of Fig.~\ref{fig:scheme}a have the form dictated by
Bloch-Floquet theorem and they depend on a wavevector \textbf{k}
which is usually restricted to the first Brillouin zone of the 2D
lattice: indeed, the effect of the dielectric modulation
$\epsilon(\textbf{r})$ is to fold the guided modes of the
effective waveguide to the first Brillouin zone and to produce
photonic bands and band gaps. We therefore write
$\textbf{g}=\textbf{k}+\textbf{G}$, where the Bloch vector
\textbf{k} lies in the first Brillouin zone and \textbf{G} is a
reciprocal lattice vector. As the basis states for the expansion
(\ref{eq:expansion}) of the magnetic field, we choose the guided
modes
$\textbf{H}_{\textbf{k}+\textbf{G},\alpha}^{\mathrm{guided}}(\textbf{r})$
of the effective waveguide given in Appendix
\ref{app:guided-modes}. The guided-mode expansion of the magnetic
field therefore reads\cite{note:alpha}
\begin{small}
\begin{equation}
\textbf{H}_{\textbf{k}}(\textbf{r})=\sum_{\textbf{G},\alpha}
c({\textbf{k}+\textbf{G},\alpha})
\textbf{H}_{\textbf{k}+\textbf{G},\alpha}^{\mathrm{guided}}(\textbf{r}).
\label{eq:gme-expansion}
\end{equation}
\end{small}
The basis set is orthonormal according to Eq.
(\ref{eq:orthonorm}), but not complete, since the radiative modes
of the effective waveguide are not included in the basis set. This
approximation will be partially lifted in Sec.~\ref{sec:losses},
where the effect of radiative modes in determining diffraction
losses will be taken into account.

With this choice for the basis set, the general index $\mu$ can be
written as $\mu\equiv(\textbf{k}+\textbf{G},\alpha)$. The matrix
elements ${\cal{H}}_{\mu\nu}$ of Eq. (\ref{eq:matrix-elements})
can be calculated in a straightforward way from the field profiles
of the guided modes: the resulting analytic expressions are
somewhat lengthy and are given in Appendix
\ref{app:guided-elements}. They depend on the inverse dielectric
matrices
\begin{small}
\begin{equation}
\eta_j(\textbf{G},\textbf{G}')=\frac{1}{A}\int_{\mathrm{cell}}
\epsilon_j(\mbox{\boldmath$\rho$})^{-1}e^{i(\textbf{G}'-\textbf{G})
\cdot\mbox{\boldmath\small$\rho$}}\,\mathrm{d}\mbox{\boldmath$\rho$}
\label{eq:eta}
\end{equation}
\end{small}
in the various layers. Like in usual 2D plane wave calculations, a
numerically convenient approach to calculate these matrix elements
is to define the dielectric matrix
\begin{small}
\begin{equation}
\epsilon_j(\textbf{G},\textbf{G}')=\frac{1}{A}\int_{\mathrm{cell}}
\epsilon_j(\mbox{\boldmath$\rho$})e^{i(\textbf{G}'-\textbf{G})
\cdot\mbox{\boldmath\small$\rho$}}\,\mathrm{d}\mbox{\boldmath$\rho$}
\label{eq:epsfour}
\end{equation}
\end{small}
and to find $\eta_j(\textbf{G},\textbf{G}')$ by numerical matrix
inversion as
$\eta_j(\textbf{G},\textbf{G}')=\epsilon_j^{-1}(\textbf{G},\textbf{G}')$:
this procedure is known to have much better convergence properties
as a function of the number of plane waves,\cite{ho90,lalanne96}
since the truncation rules for Fourier series in the presence of
discontinuous functions are better represented.\cite{li96} If the
photonic lattices in layers 1--3 have a center of inversion (the
same for all layers), the dielectric matrices
$\epsilon_j(\textbf{G},\textbf{G}')$,
$\eta_j(\textbf{G},\textbf{G}')$ are symmetric in
$\textbf{G},\textbf{G}'$ and ${\cal{H}}_{\mu\nu}$ turns out to be
a real, symmetric matrix when the phases of the fields are chosen
as in Appendix \ref{app:guided-modes}. In the most general case of
a photonic lattice without a center of inversion (like the
triangular lattice with triangular holes investigated in this
work), the matrix ${\cal{H}}_{\mu\nu}$ is complex hermitian. Line
and point defects can be treated by introducing a supercell in one
or two directions, respectively, like in 2D plane-wave
calculations.

The formalism presented here refers to the most general case of an
asymmetric PhC slab. If the structure is symmetric under
reflection by a horizontal mirror plane $z=0$ [i.e., if the upper
and lower claddings have the same dielectric pattern and are made
of the same dielectric materials,
$\epsilon_1(\mbox{\boldmath$\rho$})\equiv\epsilon_3
(\mbox{\boldmath$\rho$})$],
the basis states as well as the photonic eigenmodes can be
classified as even or odd with respect to the mirror plane. We
denote by $\hat{\sigma}_{xy}$ the corresponding specular
reflection operator and we shall refer to even (odd) states with
respect to this operator as $\sigma_{xy}=+1$ ($\sigma_{xy}=-1$)
modes. The simplest way to exploit this mirror symmetry is to
solve the eigenvalue problem (\ref{eq:linear}) separately for
$\sigma_{xy}=+1$ and $\sigma_{xy}=-1$ states by keeping only the
appropriate solutions of Eqs. (\ref{eq:tepol}),(\ref{eq:tmpol}),
as detailed in Appendix \ref{app:symmetry}. There we also discuss
the issue of polarization mixing in the framework of the GME
method: in general, a photonic mode of a PhC slab is a
superposition of TE-polarized and TM-polarized basis states and
all six components of the electric and magnetic fields are usually
non-vanishing. Another useful symmetry property holds when the
Bloch vector \textbf{k} lies along special symmetry directions,
for which mirror reflection $\hat{\sigma}_{\textbf{k}z}$ with
respect to a vertical plane ($\hat{k},\hat{z}$) is a symmetry
operation of the PhC slab: in this case the eigenmodes can be
classified as even or odd with respect to this mirror symmetry,
i.e., they have $\sigma_{\textbf{k}z}=+1$ or
$\sigma_{\textbf{k}z}=-1$ respectively. Vertical mirror symmetry
can be exploited by transforming the matrix ${\cal{H}}_{\mu\nu}$
with a unitary transformation which decouples the blocks
corresponding to $\sigma_{\textbf{k}z}=\pm 1$ states.

\subsection{Formalism for intrinsic losses}
\label{sec:losses}

When a photonic mode in the PhC slab falls above the cladding
light line (or light lines, if the waveguide is asymmetric), it is
coupled to leaky modes of the slab by the dielectric modulation
and it becomes quasi-guided, i.e., it is subject to intrinsic
losses due to scattering out of the plane. The losses can be
represented by an imaginary part of the frequency
$\mathrm{Im}(\omega)$, which is related to the Q-factor of a
resonance by $Q=\omega/(2\mathrm{Im}(\omega))$.

Within the GME method, the imaginary part of frequency can be
calculated by time-dependent perturbation theory, like in Fermi
Golden Rule for quantum mechanics. The imaginary part of the
squared frequency of a PhC mode with Bloch vector \textbf{k},
whose frequency lies above the cladding light lines (or at least
above one of them), is given by
\begin{small}
\begin{equation}
 -\mathrm{Im}\left(\frac{\omega_k^2}{c^2}\right)=\pi
 \sum_{\textbf{G}'} \sum_{\lambda=\mathrm{TE},\mathrm{TM}} \sum_{j=1,3}
 \left| {\cal{H}}_{\mathbf{k},\mathrm{rad}} \right|^2
 \, \rho_{j}\left({\bf k}+\textbf{G}';\frac{\omega_k^2}{c^2}\right),
 \label{eq:fermi}
\end{equation}
\end{small}
where the matrix element between a guided and a leaky PhC slab
mode is given by
\begin{small}
\begin{equation}
 {\cal{H}}_{\mathbf{k},\mathrm{rad}} =
\int \frac{1}{\epsilon({\bf r})}
 (\mbox{\boldmath$\nabla$}\times{\bf H}_{\textbf{k}}^*({\bf r}))\cdot
 (\mbox{\boldmath$\nabla$}\times{\bf H}_{\textbf{k}+\textbf{G}',
 \lambda,j}^{\mathrm{rad}}({\bf r}))
 \,\mathrm{d}\textbf{r} \label{eq:k-rad}
\end{equation}
\end{small}
and $\rho_{j}({\bf k}+\textbf{G}';\omega_k^2/c^2)$ is the 1D
photonic density of states (DOS) at fixed in-plane wave vector for
radiation states that are outgoing in medium $j$:
\begin{small}
\begin{eqnarray}
 \rho_{j}\left({\bf g};\frac{\omega^2}{c^2}\right)&& \equiv\int_0^{\infty}
 \frac{\mathrm{d}k_z}{2\pi} \,
 \delta\left(\frac{\omega^2}{c^2}-\frac{g^2+k_z^2}{\bar{\epsilon}_{j}}\right)
 = \nonumber \\
 &=& \frac{\bar{\epsilon}_{j}^{1/2}c}{4\pi}
 \frac{\theta\left(\omega^2-\frac{c^2g^2}{\bar{\epsilon}_{j}}\right)^{1/2}}
 {\left(\omega^2-\frac{c^2g^2}{\bar{\epsilon}_{j}}\right)^{1/2}}\,
 \label{eq:dos} .
\end{eqnarray}
\end{small}
Notice the sum over reciprocal lattice vectors and polarizations
in Eq.~(\ref{eq:fermi}), as all diffraction processes contribute
to $\mathrm{Im}(\omega^2/c^2)$.
Equations~(\ref{eq:fermi})--(\ref{eq:dos}) generalize the
expressions given in Ref.~\onlinecite{ochiai01-pert} to the case
of an asymmetric PhC slab and to situations in which processes
with all $\textbf{G}'\neq0$ contribute to diffraction losses. Once
$\mathrm{Im}(\omega^2/c^2)$ is found, the imaginary part of
frequency is easily obtained as
$\mathrm{Im}(\omega)=\mathrm{Im}(\omega^2)/(2\omega)$.

\begin{figure}[t]
\begin{center}
\vspace*{1cm} \hspace*{-1cm}
\includegraphics[width=0.6\textwidth]{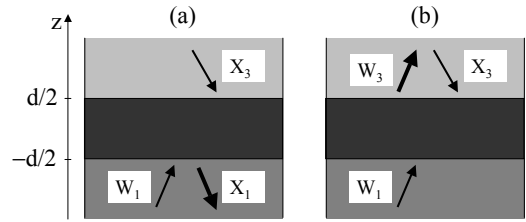}
\vspace*{-5.6cm} \caption{Schematic picture of radiative
(scattering) states for the effective waveguide with a single
outgoing component. States in (a) are outgoing in the lower
cladding (layer 1), while states in (b) are outgoing in the upper
cladding (layer 3), as indicated by the thick arrow. The notation
for the coefficients is appropriate for TE polarization, while for
TM polarization the replacements $W \rightarrow Y$, $X \rightarrow
Z$ have to be made.}\label{fig:scatt-states}
\end{center}
\end{figure}

For a given wavevector \textbf{g} and polarization $\lambda$ of the
radiative modes, there are two scattering channels, corresponding to
states with an outgoing component in the lower cladding (medium 1)
or in the upper cladding (medium 3).\cite{carniglia71,note:outgoing}
The radiation modes are schematically shown in
Fig.~\ref{fig:scatt-states}. The state which is outgoing in medium 1
corresponds to a photonic DOS with $j=1$ in Eq. (\ref{eq:dos}),
while the outgoing state in medium 3 corresponds to a DOS with
$j=3$. Radiation states are normalized according to
Eq.~(\ref{eq:orthonorm}), and only the field components in the
cladding regions are relevant in determining the
normalization.~\cite{note:normalization} Since the field profile of
a scattering state tends to a plane-wave form in the far field, the
photonic density of states (\ref{eq:dos}) is appropriate whatever
the explicit field profile of radiative PhC slab modes. In the
present GME method, we evaluate the matrix element by approximating
the radiation modes of the PhC slab with those of the effective
waveguide. This approximation is consistent with the treatment of
the previous Subsection, as the set of guided+radiation modes of the
effective waveguide is orthonormal according to
Eq.~(\ref{eq:orthonorm}), and will be shown later to yield accurate
results for diffraction losses when compared to more exact methods.
In Appendix \ref{app:rad-modes} we give the explicit form of
radiation states of the effective waveguide which are outgoing in
either the lower cladding (Fig.~\ref{fig:scatt-states}a) or in the
upper cladding (Fig.~\ref{fig:scatt-states}b), for both TE and TM
polarizations. Using expansion (\ref{eq:gme-expansion}) for the
magnetic field of a PhC slab mode, the matrix element
(\ref{eq:k-rad}) becomes
\begin{small}
\begin{equation}
 {\cal{H}}_{\mathbf{k},\mathrm{rad}} = \sum_{\textbf{G},\alpha}
c({\textbf{k}+\textbf{G},\alpha})^* \,
{\cal{H}}_{\mathrm{guided},\mathrm{rad}},
 \label{eq:k-rad-gme}
\end{equation}
\end{small}
where
\begin{small}
\begin{equation}
\, {\cal{H}}_{\mathrm{guided},\mathrm{rad}} = \int
\frac{1}{\epsilon({\bf r})} (\mbox{\boldmath$\nabla$}
\times[\textbf{H}_{\textbf{k}+\textbf{G},\alpha}^{\mathrm{guided}}
(\textbf{r})^*])\cdot
(\mbox{\boldmath$\nabla$}\times\textbf{H}_{\textbf{k}+\textbf{G}',
\lambda,j}^{\mathrm{rad}}({\bf r})) \, \mathrm{d}\textbf{r}
 \label{eq:guided-rad-gme}
\end{equation}
\end{small}
are matrix elements between guided and radiation modes
of the effective waveguide, which can be calculated analytically.
In Appendix \ref{app:rad-elements} we collect the matrix elements
(\ref{eq:guided-rad-gme}), for all possible combinations of
polarizations. Those expressions are implemented together with
Eqs.~(\ref{eq:fermi})--(\ref{eq:guided-rad-gme}), in order to
evaluate the imaginary part of frequency and hence the intrinsic
diffraction losses of quasi-guided modes.

The calculation of losses is computationally more demanding than
the photonic dispersion alone, as the eigenvectors of the linear
problem (\ref{eq:linear}) are also needed. The formalism for
intrinsic diffraction losses presented in this Subsection and in
the Appendix refers to the general case of an asymmetric PhC slab.
If the slab is symmetric, and if the PhC slab modes are found as
$\sigma_{xy}=\pm1$ states as outlined in Appendix
\ref{app:symmetry}, it is possible to define and use symmetric or
antisymmetric outgoing states instead of the scattering states
illustrated in Fig.~\ref{fig:scatt-states}. If this is done,
$\sigma_{xy}=+1$ ($\sigma_{xy}=-1$) quasi-guided modes are coupled
only to symmetric (antisymmetric) radiative states. From a
computational point of view there is no special advantage in using
symmetry-adapted radiative states, as the computing time needed to
implement the perturbative formula Eq.~(\ref{eq:fermi}) is usually
small. Thus even in the case of a symmetric PhC slab it is
convenient to use the general loss formalism, and simply obtain
$\sigma_{xy}=+1$ or $\sigma_{xy}=-1$ PhC slab states by selecting
the basis states of the effective waveguide, as explained in
Appendix \ref{app:symmetry}.

\subsection{Discussion of the method}
\label{sec:convergence}

In this Subsection we discuss the convergence properties of the
GME as a function of numerical parameters and give a brief
overview of previously published results and comparison with other
approaches. Truncation parameters in the GME method are the choice
and number $N_{\mathrm{PW}}$ of plane waves and the number
$N_{\alpha}$ of guided modes of the effective waveguide kept in
the expansion. The dimension of the eigenvalue problem
(\ref{eq:linear}) is $N_{\mathrm{PW}}\cdot N_{\alpha}$, or about
half of this value when vertical parity
$\hat{\sigma}_{\textbf{k}z}$ is used. The choice of plane waves
can be made in analogy to usual 2D plane-wave expansion: the
important point is how to calculate the inverse dielectric matrix
$\eta_j(\textbf{G},\textbf{G}')$ properly, as already noticed in
Subsec.~\ref{sec:dispersion}. In the present implementation we
take an isotropic cutoff for reciprocal lattice vectors and adopt
the common procedure of inverting the dielectric matrix
numerically:\cite{ho90} while this is adequate for many purposes,
improved implementations of the plane-wave expansion
\cite{lalanne98,david06} could also be introduced in the GME
method in order to speed up convergence.

\begin{figure}[t]
\includegraphics[width=0.38\textwidth]{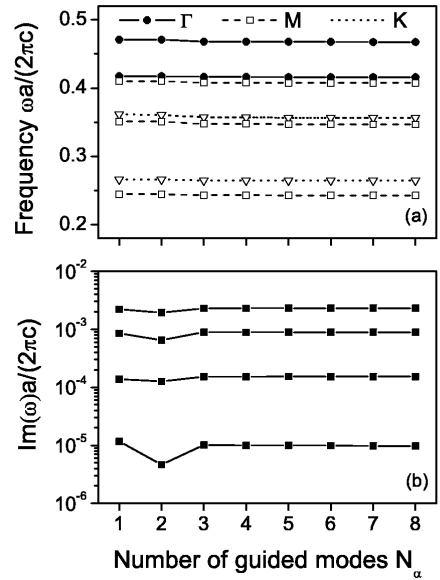}
\vspace{-0.5cm} \caption{(a) Real part and (b) imaginary part of
frequency as a function of number of guided modes $N_{\alpha}$ in
the expansion, for $\sigma_{xy}=+1$ modes in the triangular
lattice of circular air holes defined in a high-index suspended
membrane. The frequencies in (a) are calculated at the symmetry
points $\Gamma$,K,M, while the imaginary parts in (b) are
calculated at $k=\pi a/3$ along the $\Gamma$-K direction.
Parameters are: $r/a=0.3$, $d/a=0.5$, $\varepsilon=12.11$. The
effective core dielectric constant is fixed to
$\bar{\varepsilon}_2=8.4827$, as given by Eq.~(\ref{eq:average}).
} \label{fig:conv_nal}
\end{figure}

The optimal $N_{\alpha}$ depends on waveguide parameters and on
the frequency region considered: typically, for PhC slabs defined
in a thin high-index membrane, $N_{\alpha}=4$ in each parity
sector (i.e., for both $\sigma_{xy}=+1$ and $\sigma_{xy}=-1$
modes) yields very accurate results for the lower-lying modes. As
an example, in Fig.~\ref{fig:conv_nal} we show the real and
imaginary parts of the frequency for a few photonic modes as a
function of the number of guided modes $N_{\alpha}$, for
$\sigma_{xy}=+1$ modes in the triangular lattice of circular air
holes. Structure parameters are similar to those used later in the
paper. The number of plane waves is truncated to
$N_{\mathrm{PW}}=109$, which is sufficient for convergence in the
present lattice. It can be seen that both the mode energies and
the imaginary parts are stable for $N_{\alpha}\ge 3$, while a good
approximation is obtained already for $N_{\alpha}=2$ or even
$N_{\alpha}=1$. This is very convenient for the purpose of design
and search of parameters, as a photonic structure in a high-index
contrast PhC slab can be optimized within a fully 3D approach,
with the same computing time of a conventional 2D plane-wave
expansion. The precise calculation (e.g., with $N_{\alpha}\ge4$)
can be performed \emph{a posteriori} once the structure parameters
are optimized.

\begin{figure}[t]
\includegraphics[width=0.38\textwidth]{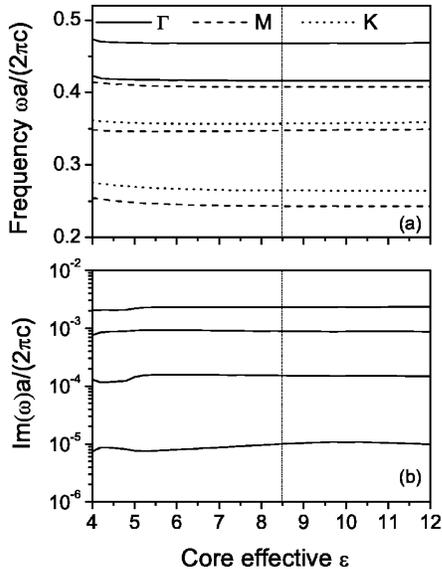}
\vspace{-0.5cm} \caption{(a) Real part and (b) imaginary part of
frequency as a function of core dielectric constant in the
effective waveguide, for $\sigma_{xy}=+1$ modes in the triangular
lattice of circular air holes defined in a high-index suspended
membrane. The frequencies in (a) are calculated at the symmetry
points $\Gamma$,K,M, while the imaginary parts in (b) are
calculated at $k=\pi a/3$ along the $\Gamma$-K direction.
Parameters are: $r/a=0.3$, $d/a=0.5$, $\varepsilon=12.11$. The
number of guided modes in the expansion is fixed to
$N_{\alpha}=4$. The vertical bar indicates the average dielectric
constant, Eq.~(\ref{eq:average}), usually employed in the GME
method (in the present case, $\bar{\varepsilon}_2=8.4827$).
}\label{fig:conv_eps}
\end{figure}

It is also important to discuss the choice of the dielectric
constant of the effective waveguide, which is usually taken to be
the spatial average (\ref{eq:average}) in each layer. While this
choice is a natural one, it might be asked whether the results
depend on the assumed value of $\bar{\epsilon}_j$. In
Fig.~\ref{fig:conv_eps} we show the real and imaginary parts of
frequency for a few modes, with the same parameters of the
previous Figure, as a function of the effective dielectric
constant in the core region. It can be seen that the results are
rather insensitive to the assumed value of $\bar{\epsilon}_2$,
except when it becomes smaller than about five. All curves are
flat around the value $\bar{\epsilon}_2=8.4827$ which corresponds
to the spatial average. This satisfactory behavior arises because
the GME method employs the set of \emph{all} guided modes of the
effective waveguide (although in practice they are truncated to a
finite number for numerical convenience), thereby compensating for
different choices of the effective waveguide used to define the
basis set. Thus the choice of the dielectric constant of the
effective waveguide as the spatial average of
$\bar{\epsilon}_j(\textbf{r})$ is justified, at least for the low
air fractions that are commonly employed. Although one might
consider extending the method by introducing an anisotropic
dielectric tensor for each layer of the effective waveguide,
Fig.~\ref{fig:conv_eps} strongly suggests that the results of the
GME method would not be changed by such a complicated extension
which is therefore unnecessary.

The approximations of the GME method are: (i) for photonic
dispersion, the shift of guided and quasi-guided modes due to
second-order coupling to leaky modes of the effective waveguide is
neglected; (ii) for diffraction losses, the density of radiative
modes of the PhC slab is approximated with that of the effective
waveguide. The effect of these approximations can be judged by
comparing the results with those of other methods that are known
to be exact within numerical accuracy. A full analysis of this
issue is outside the scope of this paper, however comparisons
between the results of GME and those obtained by other methods
have already been published and are shortly summarized here. The
dispersion of quasi-guided modes was found to be in very good
agreement with the resonance positions in reflectance or
transmittance spectra found from a scattering-matrix
treatment.\cite{andreani02_ieee} In fact, the complex frequencies
of photonic modes can be obtained from the poles of a scattering
matrix.\cite{lalanne02,tikhodeev02,morand03} The intrinsic
propagation losses of line-defect waveguides in the triangular
lattice obtained with different methods (GME,\cite{andreani03_apl}
finite-difference time-domain or FDTD\cite{cryan05} and from the
poles of a scattering matrix\cite{lalanne02,sauvan05_thesis}) were
compared for the same waveguide parameters and were found to agree
very well with each other: a detailed comparison is presented in
Ref.~\onlinecite{sauvan05_thesis}. A comparative study of GME and
Fourier modal methods applied to cavity modes in one-dimensional
PhC slabs has been reported in Ref.~\onlinecite{gerace05_oqe}, and
very good agreement was found for the photonic dispersion as well
as the intrinsic losses. Finally, the Q-factors of cavity modes in
L1, L2 and L3 nanocavities (one, two or three missing holes in the
triangular lattice) as a function of nearby hole displacement are
found to be almost identical when calculated with either the GME
method\cite{andreani04_pn} or the poles of a scattering
matrix.\cite{sauvan05_prb} Agreement with experiments performed on
PhC slabs made of
Silicon-on-Insulator,\cite{patrini02,galli04,galli05_jsac} Si
membranes,\cite{galli05_prb}
GaAs/AlGaAs,\cite{galli02_epjb,malvezzi03} and Silicon
Nitride\cite{gerace05_apl} is also very satisfactory. The set of
these comparisons indicates that the GME is a reliable method for
both photonic mode dispersion and intrinsic losses, whose main
advantages are computational efficiency and ease of application to
various kinds of vertical waveguides and of photonic lattices,
both periodic in 2D and containing line- and point defects.

\begin{figure}[t]
\includegraphics[width=0.4\textwidth]{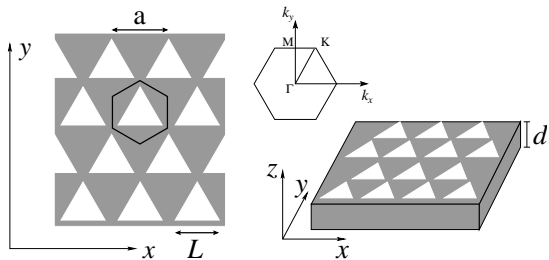}
\caption{Schematic picture of the triangular lattice of triangular
air holes with lattice constant $a$ and triangle side $L$; the
unit cell in real space is indicated. The lattice is patterned in
a high index suspended membrane of thickness $d$. The Brillouin
zone with main symmetry directions in reciprocal space is also
drawn.} \label{fig:scheme-tria}
\end{figure}

\section{Triangular lattice with triangular holes}
\label{sec:trianglatt}

We report in this Section a systematic study of the triangular
lattice of triangular air holes patterned on a free-standing
high-index slab of thickness $d$, as schematically shown in
Fig.~\ref{fig:scheme-tria}. The relevant dimensions of this
lattice are the triangular basis side, $L/a$, and the slab
thickness. Throughout this Section we assume (unless otherwise
specified) that the lattice is patterned in a material with
dielectric constant $\varepsilon=12.11$, which is appropriate for
Silicon at the usual telecom wavelength $\lambda=1.55$ $\mu$m but
also to other materials with similar dielectric constants (e.g.,
GaAs). For the calculations shown in the present Section, we
employed up to 109 plane waves and 4 guided modes per parity
sector in the basis set of the GME method. All the calculations
have been done on an ordinary Pentium computer with low
computational effort. The complex Fourier transform of the
dielectric constant for the triangular lattice of triangular holes
is given in Appendix~\ref{app:four-trans}. Notice that the lattice
is invariant under rotations by $120^{\circ}$ and is not
centrosymmetric, but since
$\omega(-\textbf{k})=\omega(\textbf{k})$ by time-reversal
symmetry, the same symmetry directions of the usual triangular
lattice of circular air holes do apply.

\begin{figure}[t]
\includegraphics[width=0.4\textwidth]{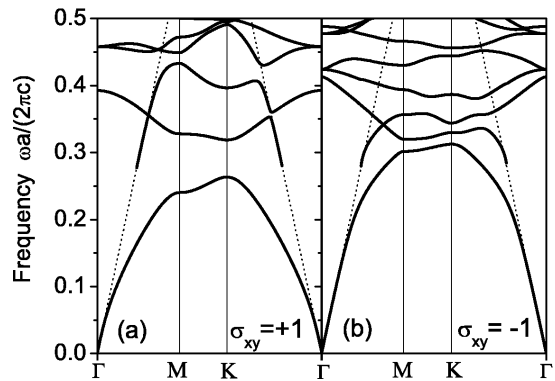}
\caption{Photonic band dispersion for (a) even ($\sigma_{xy}=+1$)
and (b) odd ($\sigma_{xy}=-1$) modes of the PhC slab with a
triangular lattice of triangular holes, parameters as in Ref.
\onlinecite{takayama05}: $L/a=0.85$, $d/a=0.68$,
$\varepsilon=12.2$. Light dispersion in air is represented by
dotted lines.}\label{fig:takayama}
\end{figure}

\begin{figure*}[t]
\includegraphics[width=0.85\textwidth]{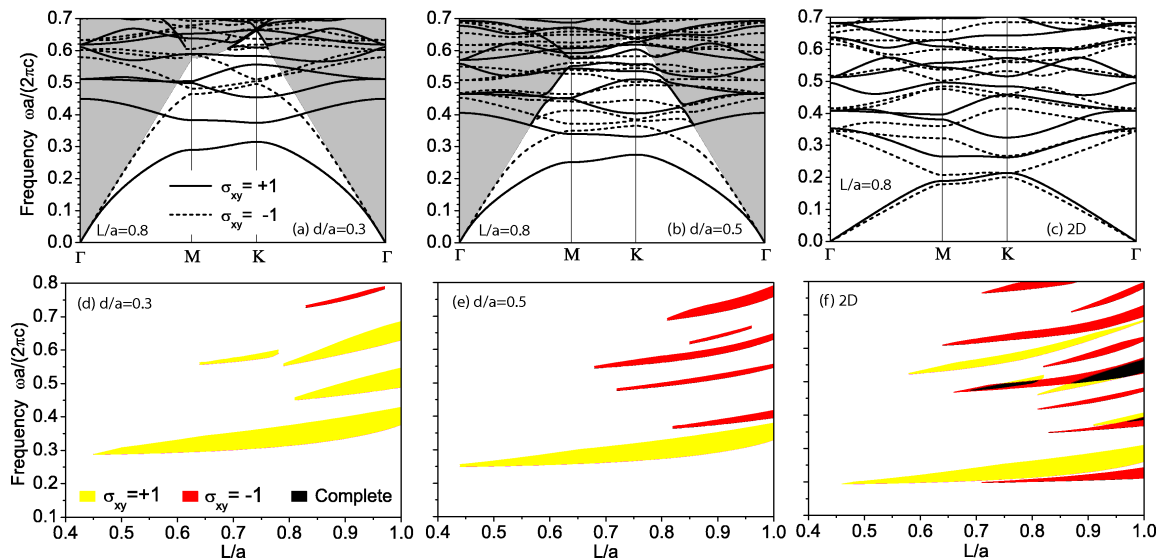}
\caption{(Color online) Upper panels: photonic band dispersion for
$L/a=0.8$ in (a) PhC slab of thickness $d/a=0.3$, (b) $d/a=0.5$,
and (c) ideal 2D photonic crystal. Lower panels: gap maps as a
function of the holes' side for (d) $d/a=0.3$, (e) $d/a=0.5$, and
(f) ideal 2D system. The high index material is assumed in all of
these calculations to have a dielectric constant
$\varepsilon=12.11$. Light grey regions represent the modes above
the light line for the waveguide-embedded systems.
}\label{fig:gapmaps}
\end{figure*}

The first realization of such a lattice has been reported in
Ref.~\onlinecite{takayama05}, with the purpose of measuring a
complete photonic band gap in a waveguide-embedded 2D PhC thanks
to a reduction of symmetry for the basis in the unit cell of the
triangular lattice, as already pointed out in the Introduction.
The importance of finding a complete photonic band gap in a
waveguide-embedded photonic structure is related to the practical
realization of linear waveguides or cavities with reduced losses
or higher Q-factors, as scattering into quasi-guided modes of
opposite parity would be avoided. In Fig.~\ref{fig:takayama} we
present the photonic band dispersion calculated with the GME
method for structure parameters as in
Ref.~\onlinecite{takayama05}. Even (TE-like) and odd (TM-like)
modes are reported in Figs.~\ref{fig:takayama}a and b,
respectively. It is worth noticing that in the absence of
second-order waveguide modes, the fundamental gaps of both
parities would overlap in the dimensionless frequency range
0.31-0.32, as it can be seen from the Figure. However, the
presence of a second-order mode with cut-off around $\omega
a/(2\pi c)\simeq 0.28$ for both polarizations, as shown in
Fig.~\ref{fig:takayama}, prevents the opening of a true photonic
band gap. Such second-order modes (which do not appear in a 2D
treatment with an effective index) are unavoidable in the present
structure, due to the relatively large slab thickness, $d/a\simeq
0.7$.

We believe that our result is not in contrast with the
experimental findings reported in Ref.~\onlinecite{takayama05}, in
which a strong reduction of the transmission intensity was
measured along the $\Gamma$K and $\Gamma$M directions. Since the
dispersion of the second-order waveguide mode is very close to the
air light line in the relevant frequency region, such modes are
barely coupled to the external beam focused on the sample side and
also they are subject to large disorder-induced losses; thus, it
is likely that transmission into such modes is attenuated in a few
lattice periods, which would explain the apparent photonic band
gap observed experimentally. On the other hand, when line- or
point-defects are introduced, such second-order modes are
necessarily present within the gap, thus limiting the usefulness
of this design for applications. In the following, we will
consider a photonic band gap for the waveguide-embedded structure
only when no higher-order modes are present within the apparent
gaps formed by the fundamental modes, i.e., when the PhC slab is
truly single-mode.

With the aim of optimizing the design proposed by Takayama
\textit{et al.}, we performed a systematic study of the present
photonic lattice at varying structure parameters. In
Fig.~\ref{fig:gapmaps} we show the photonic band dispersion and
gap maps for the triangular lattice of triangular air holes with
slab thickness $d/a=0.3$ (Figs.~\ref{fig:gapmaps}a and d),
$d/a=0.5$ (Figs.~\ref{fig:gapmaps}b and e), and for the reference
2D system of infinite thickness (Figs.~\ref{fig:gapmaps}c and f),
respectively. As is well known for more conventional lattices,
\cite{johnson99,andreani02_ieee} the blue shift of photonic modes
with respect to the ideal 2D system due to the dielectric
confinement is strongly polarization-dependent; in particular, it
is more pronounced for odd (TM-like) modes. This physical behavior
is seen by comparing Figs.~\ref{fig:gapmaps}a-c, in which the
dispersion of photonic modes is calculated for the same 2D
photonic lattice in which triangular air holes have side
$L/a=0.8$. For $d/a=0.3$ the second-order cut-off is rather high
in energy, thus the system can be considered as single-mode below
the light line. On increasing the slab thickness, the second-order
cut-off red shifts. For $d/a=0.5$, both even and odd modes open
their fundamental band gaps, but they do not overlap to give a
complete photonic band gap for all symmetry directions. In order
to span a wider range of structure parameters, important
information can be inferred by the gap width as a function of the
holes' side. As it can be seen in Figs.~\ref{fig:gapmaps}d-f, the
gap maps depend strongly on the waveguide thickness and differ
substantially from those of the 2D system. For $d/a=0.3$, the
fundamental odd gap is closed for all values of $L/a$, due to the
larger blue shift of band edges at the K-point than at the
M-point. Even if both fundamental band gaps are present for
$d/a=0.5$, no complete gap opens due to the stronger blue shift of
odd modes. For $d/a>0.5$, the presence of higher-order modes
prevents the opening of complete gaps as well. The gap map for the
ideal 2D system presents a great variety of photonic gaps for both
polarizations, and also some complete gaps at high frequencies. It
is worth noting that the fundamental TE and TM band gaps do not
overlap in frequency, unlike in the ordinary triangular lattice of
circular air holes.\cite{joanno} Furthermore, the fundamental TE
gap turns out to be narrower than the usual gap for circular
holes.

\begin{figure}[t]
\includegraphics[width=0.4\textwidth]{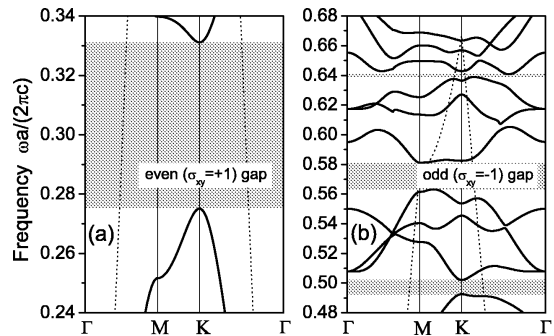}
\caption{Photonic band dispersion for (a) $\sigma_{xy}=+1$ modes
at pump and (b) $\sigma_{xy}=-1$ modes at second-harmonic
frequencies, respectively, showing resonant band gaps at $\omega$
and $2\omega$ useful for non-linear applications. Parameters are
as in Fig. \ref{fig:gapmaps}b: $L/a=0.8$, $d/a=0.5$,
$\epsilon=12.11$. Light dispersion in air is represented by dotted
lines.}\label{fig:shg}
\end{figure}

Even if no truly complete band gaps appear to be present for this
waveguide-embedded lattice, it is interesting to investigate the
formation of resonant gaps at $\omega$ and $2\omega$, to be
exploited in non-linear optical applications like second-harmonic
generation (SHG). While doubly-resonant SHG has been widely
studied in one-dimensional systems like multilayers and
microcavities,\cite{berger97,simonneau97,centini99,deangelis01,dumeige02,liscidini04apl,
ochiai05,liscidini06pre} no 2D photonic lattice giving such an
interesting result has been reported in the literature. In
particular, it can be seen from Fig.~\ref{fig:gapmaps}e that many
odd gaps are present for a slab thickness $d/a=0.5$. In this case,
it is easy to find a condition for which the fundamental
$\sigma_{xy}=+1$ gap at pump frequency $\omega$ is resonant with a
$\sigma_{xy}=-1$ gap at $2\omega$. One of the possibilities is
indeed shown in Fig.~\ref{fig:shg}. It is substantially the same
result reported in Fig.~\ref{fig:gapmaps}b, but here the odd modes
(Fig.~\ref{fig:shg}b) are plotted in a doubled frequency range
with respect to the even modes (Fig.~\ref{fig:shg}a). As it can be
seen, there is a good overlap in frequency between the large band
gap of Fig.~\ref{fig:shg}a and the rather narrow but robust gap of
Fig.~\ref{fig:shg}b. This could be interesting, e.g., for
applications to non-linear optical converters based on GaAs, in
which the pump s-polarized field would be converted into a
p-polarized second-harmonic field.\cite{note:SHG} In this respect,
the usefulness of the present lattice would be enhanced with
\textit{ad hoc} fabricated line- or point-defects. It should be
reminded that these results have been obtained by using the same
dielectric constant for both pump and second-harmonic modes, but
it is easy to extend the present calculations to non-linear
materials with optical constants of known frequency dispersion.

\begin{figure}[t]
\includegraphics[width=0.4\textwidth]{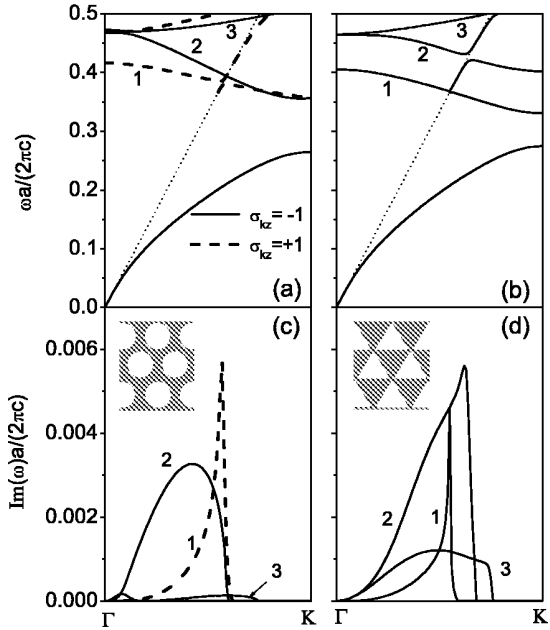}
\caption{(a) Photonic band dispersion and (c) intrinsic losses for
$\sigma_{xy}=+1$ modes in the triangular lattice of circular air
holes with $r/a=0.3$, $d/a=0.5$, $\varepsilon=12.11$, along the
$\Gamma$-K symmetry direction; modes are classified according to
their $\hat{\sigma}_{kz}$ parity. (b) Dispersion and (d) intrinsic
losses for the same lattice of triangular air holes with
$L/a=0.8$, $d/a=0.5$, $\varepsilon=12.11$. Light dispersion in air
is given by dotted lines.}\label{fig:bulk-loss}
\end{figure}

One of the issues when dealing with the triangular lattice of
triangular holes would be given by intrinsic losses. In order to
show the general interest of the triangular hole-based lattice,
and also to show an application of the GME method, we compare
intrinsic losses of the present system to the ones of an ordinary
circular hole-based PhC slab. In this case, we restrict our
discussion to $\sigma_{xy}=+1$ modes along the $\Gamma$K symmetry
direction. Intrinsic losses are quantified by calculating the
imaginary part of frequency, as detailed in the previous Section.
In Fig.~\ref{fig:bulk-loss}a and c we show the photonic dispersion
and intrinsic losses for the first few bands of an ordinary PhC
slab with a triangular lattice of circular holes. It can be
noticed that the modes along the chosen symmetry direction are
separated according to their mirror symmetry with respect to the
vertical plane, which can be either even ($\sigma_{kz}=+1$) or odd
($\sigma_{kz}=-1$).\cite{note:sakoda} On the right hand side, in
Fig.~\ref{fig:bulk-loss}b and d, we plot dispersion and losses for
the case of triangular holes. In this case no symmetry operation
$\hat{\sigma}_{\textbf{k}z}$ holds, so the modes are mixed. The
choice of $r/a=0.3$ for the circular holes and $L/a=0.8$ for
triangular ones, respectively, allows to obtain roughly the same
air fraction $f\simeq0.32$ for the two lattices. This can be also
verified by the roughly equal second-order cut-offs. The
dispersion is only slightly modified when changing the hole shape
from circular to triangular, and in particular it appears that a
general blue shift occurs for the modes in
Fig.~\ref{fig:bulk-loss}b with respect to the ones in
~\ref{fig:bulk-loss}a. For what concerns intrinsic losses, we can
see that they remain almost of the same order on going from
circular to triangular holes, at least for the lower frequency
bands. This means that (in the absence of disorder) no particular
differences are introduced by changing the basis of the lattice,
apart from the loss of symmetry $\hat{\sigma}_{\textbf{k}z}$. As
the band labelled with 1 is almost unchanged on going from the
circular to the triangular hole shape, also the intrinsic losses
are quite similar between the two lattices. For higher lying
bands, in particular for the band labelled with 2, such similarity
becomes less evident on approaching the light line, owing to the
blue shift of modes for the triangular hole-based lattice and to
the mixing with other modes which does not occur in the
conventional lattice. In particular, owing to the lack of
$\hat{\sigma}_{\textbf{k}z}$ symmetry, there is an anticrossing
between mode 2 and the second-order waveguide mode occurring very
close to the light line, which correspondingly affects the
imaginary part of frequency (see Figs.~\ref{fig:bulk-loss}b-d).
The mode labelled with 3 has larger losses for triangular holes
than for circular ones, but it is quite high in energy.

As a conclusion to this section, we have shown that the triangular
lattice of triangular air holes does indeed have a complete
photonic band gap between the fundamental waveguide modes, but the
gap is actually eliminated by the presence of second-order modes.
The lattice can be designed for nonlinear optical applications in
order to have a gap for even modes at the pump frequency which is
resonant with a gap for odd modes at the SH frequency. Intrinsic
losses are of the same order as in the circular hole-based
triangular lattice. The GME method is useful for a comparison of
the two lattices and for the design of structures with desired
characteristics.

\section{Line-defect waveguides and point cavities}
\label{sec:defects}

We show in the present Section that line- and point-defects can be
introduced into the triangular-hole lattice without any loss of
performance with respect to the circular-hole one. It is well
known that removing a row of holes in a 2D photonic lattice
introduces a linear defect supporting guided modes within the
photonic band gap. In a triangular lattice such waveguide is
usually called W1, and its channel width (namely the distance
between the holes surrounding the waveguide) is $w=w_0=\sqrt{3}a$.
These guided modes exploit the usual index confinement along the
vertical direction added to the gap confinement in the waveguide
plane, and they are studied with great interest by the scientific
community owing to prospective applications in photonic integrated
circuits.

\begin{figure}[t]
\includegraphics[width=0.4\textwidth]{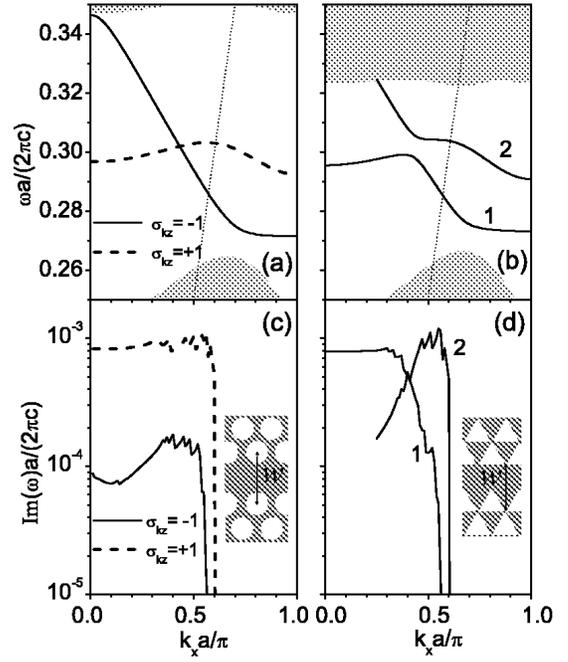}
\caption{(a) Photonic band dispersion and (c) intrinsic losses for
$\sigma_{xy}=+1$ modes in W1 waveguides in triangular lattice of
circular air holes with $r/a=0.3$, $d/a=0.5$, $\varepsilon=12.11$;
modes are classified according to their $\hat{\sigma}_{kz}$
parity. (b) Dispersion and (d) intrinsic losses for the same
waveguide in the triangular lattice of triangular air holes with
$L/a=0.8$, $d/a=0.5$, $\varepsilon=12.11$. Light dispersion in air
is given by dotted lines. The shaded regions represent the bulk
modes of the respective lattices projected onto the line-defect
Brillouin zone.}\label{fig:w1-loss}
\end{figure}

In Fig.~\ref{fig:w1-loss} we report a comparison between the W1
waveguide realized in the triangular lattice with circular
(Figs.~\ref{fig:w1-loss}a-c) and triangular holes
(Figs.~\ref{fig:w1-loss}b-d), respectively. For the circular
hole-lattice the guided modes are classified according to their
mirror symmetry with respect to the vertical plane $kz$. For such
calculations, we used supercells along the $\Gamma$M direction of
widths ranging from $4w_0$ to $9w_0$, with 181 to 323 plane waves
and 4 guided modes in the basis set for the expansion. Then, an
average was taken to get final results for losses, in order to
smooth out finite supercell effects.\cite{andreani03_apl} As it is
seen from the Figure, the dispersion of the guided modes is very
similar between the two systems, apart in the region of mixing. In
fact, as $\hat{\sigma}_{\textbf{k}z}$ is a symmetry operation for
the system in Fig.~\ref{fig:w1-loss}a, the two guided modes are
completely decoupled, while in Fig.~\ref{fig:w1-loss}b they mix
producing a consistent anti-crossing and a mini-gap. Looking at
the imaginary part of frequency, we see by comparing
Figs.~\ref{fig:w1-loss}c and d that they are almost identical away
from the mixing region, which means that W1 waveguides in
triangular lattice of triangular holes may behave as standard W1
waveguides.\cite{note:proploss} Such result is quite surprising,
as the asymmetry of such a waveguide could be expected to lead to
higher scattering of propagating light states. On the contrary,
the reduction of symmetry does not increase losses in the
\emph{perfect} photonic lattice - of course, the role of
fabrication imperfections may affect differently the two
structures in \emph{real} lattices.

\begin{figure*}[t]
\includegraphics[width=0.85\textwidth]{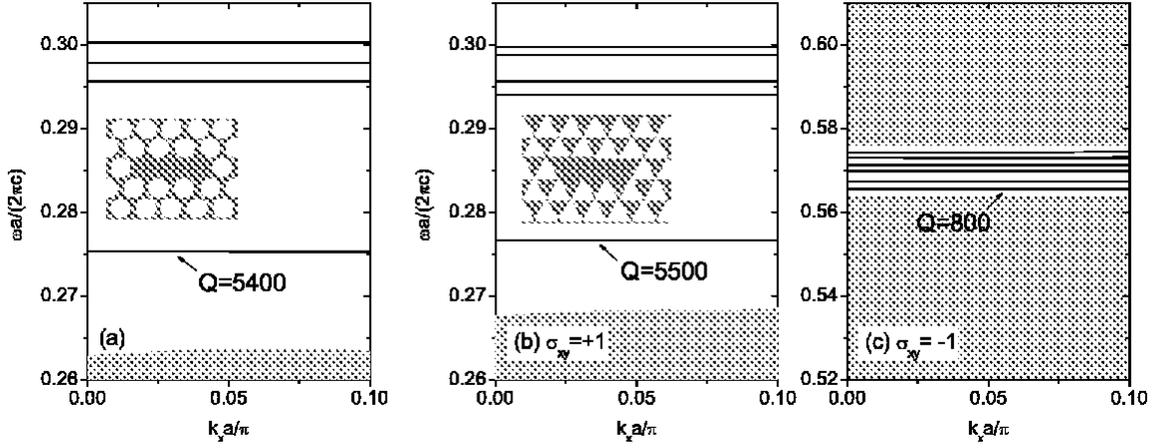}
\caption{Cavity modes for (a) L3 cavity in triangular lattice of
circular holes, $\sigma_{xy}=+1$ modes, with $r/a=0.3$, $d/a=0.5$,
$\varepsilon=12.11$; (b) L3 cavity in triangular lattice of
triangular holes, $\sigma_{xy}=+1$ modes at pump frequency, with
$L/a=0.8$, $d/a=0.5$, $\varepsilon=12.11$; (c) L3 cavity in
triangular lattice of triangular holes, $\sigma_{xy}=-1$ modes at
second harmonic frequency. Shaded regions represent the bulk modes
of the respective lattices projected onto the point-defect
Brillouin zone.}\label{fig:cavity}
\end{figure*}

The final goal of this paper is to exploit the potential of the
triangular lattice of triangular holes for non-linear optical
applications. In this respect, the ultimate system for frequency
conversion at the nanoscale is a photonic cavity with double
resonance for pump and harmonic waves. Doubly-resonant
microcavities for second-harmonic generation have been studied in
the context of Fabry-Perot
cavities.\cite{berger97,simonneau97,liscidini04apl,ochiai05,liscidini06pre}
According to a recent proposal, double resonance in photonic
crystal cavities may allow to achieve strong coupling between
single photons.\cite{irvine06} It has been shown that full 3D
confinement of light with a very high figure of merit (quality
factor, Q) can be achieved by properly designed point-defects in
2D PhC slabs.\cite{akahane03} In particular, removing three holes
along the $\Gamma$K direction (L3 cavity) and optimizing the
positions of the holes surrounding the cavity allows to achieve
Q-factors of the order of $10^5$.\cite{akahane05} We compare here
the L3 cavity structure made in a triangular lattice of triangular
air holes to the usual L3 cavity with circular holes and the
results are shown in Fig.~\ref{fig:cavity}. For such calculations,
a square supercell of dimensions $10a\times 5w_0$ has been taken.
Up to 1551 plane waves and two guided modes were used in the GME.
As it can be seen from Fig.~\ref{fig:cavity}a, many dispersionless
$\sigma_{xy}=+1$ defect modes appear in the photonic band gap, due
to the tight confinement provided by the cavity. We can evaluate
the vertical Q-factor of such modes (i.e., the one determined by
intrinsic out-of-plane losses) by the definition
$Q=\omega/[2\mathrm{Im}(\omega)]$, where both $\omega$ and
$\mathrm{Im}(\omega)$ have been averaged over the folded Brillouin
zone in order to reduce the effects of a finite supercell size.
Very good agreement with measured Q-factors and defect mode
energies has been already found with this GME-based
procedure.\cite{andreani04_pn} The Q-factor of the un-optimized L3
cavity in the lattice of circular holes is theoretically estimated
to be $Q\simeq 5400$. When the same cavity is realized in the
triangular lattice of triangular holes (Fig.~\ref{fig:cavity}b),
we see that the fundamental mode is almost unchanged, and its
Q-factor is comparable to the one of Fig.~\ref{fig:cavity}a. The
slight blue shift of the defect mode and of the lower band edge in
Fig.~\ref{fig:cavity}b with respect to Fig.~\ref{fig:cavity}a is
due to the slightly different air fractions of the two lattices.
As we have already shown in Fig.~\ref{fig:shg}, we can design a
triangular lattice with triangular holes having a doubly-resonant
band gap. Thus, we plot in Fig.~\ref{fig:cavity}c the dispersion
of $\sigma_{xy}=-1$ modes within the doubled frequency range with
respect to Fig.~\ref{fig:cavity}b, in analogy with
Fig.~\ref{fig:shg}. Many defect modes appear within the $2\omega$
band gap, and the lowest mode has a Q-factor of almost $10^3$,
which is a very promising value considering that the structure has
not been optimized in any way. There is, in fact, much room for
improving the Q-factors at both $\omega$ and $2\omega$ by means of
geometry optimization of the nearby holes. The present result make
us confident that high-Q, doubly-resonant nanocavities may be
realized in PhC slabs made of non-linear materials, bringing new
exciting results in integrated optics at nanoscale level.

\section{Conclusions}

The triangular lattice of triangular holes in a high-index
dielectric slab has been thoroughly investigated by means of the
guided-mode expansion method, whose detailed formulation and
examples of applications have been given in the paper. The
investigated lattice does have a complete band gap for both
polarizations when the fundamental waveguide mode is considered,
consistently with the results of Ref.~\onlinecite{takayama05}, but
the gap is actually eliminated by the presence of second-order
waveguide modes. The intrinsic losses of quasi-guided modes in the
triangular lattice with triangular and circular holes, and in
line-defect waveguides as well as nanocavities defined therein,
are comparable.

An interesting application of the triangular lattice of triangular
holes for nonlinear optics follows from the fact that a gap for
$\sigma_{xy}=+1$ (TE-like) modes at the pump frequency can be
designed to be in resonance with a gap for $\sigma_{xy}=-1$
(TM-like) modes at the harmonic frequency: thus, resonant
second-harmonic generation exploiting band-edge effects may be
realized. Promising results are also found for point cavities: the
L3 cavity in the triangular lattice of triangular holes supports a
$\sigma_{xy}=+1$ cavity mode at $\omega$ which can be resonant
with a $\sigma_{xy}=-1$ cavity mode at $2\omega$. There is much
room for improving the design of doubly-resonant, high-Q
nanocavities by geometry optimization. These results are
interesting in view of application of the triangular lattice of
triangular holes for nanoscale nonlinear optical processes.

\section*{Acknowledgments}
The authors are grateful to M.~Agio, H.~Benisty, D.~Cassagne,
Ph.~Lalanne, M.~Le~Vassor d'Yerville, M.~Liscidini, and C.~Sauvan
for useful conversations and suggestions. They also would like to
acknowledge K.~Hennessy and W.T.M.~Irvine for providing them with
the preprint of Ref.~\onlinecite{irvine06} prior to its
publication, thereby motivating the search for doubly resonant PhC
cavities. This work was supported by MIUR through Cofin project
``Silicon-based photonic crystals for the control of light
propagation and emission'' and FIRB project ``Miniaturized
electron and photon systems'', as well as by CNR-INFM through PRA
PHOTONIC.

\appendix


\section{Guided modes of the effective waveguide}
\label{app:guided-modes}

The frequencies of the guided modes of the effective waveguide are
found from the implicit equations (\ref{eq:tepol}) for TE
polarization, and (\ref{eq:tmpol}) for TM polarization. The guided
modes are labelled by the wavevector \textbf{g} and the mode index
$\alpha$,\cite{note:alpha} which can be combined into a single
index $\mu=(\textbf{g},\alpha)$. The mode profiles for TE
polarization can be written as follows (with the time dependence
$e^{-i\omega_{\mu}t}$ being understood), with reference to the
geometry of Fig.~\ref{fig:scheme}:
\begin{small}
\begin{widetext}
\begin{equation}
\textbf{E}_{\mu}^{\mathrm{guided}}(\mbox{\boldmath$\rho$},z)=\frac
{e^{i\textbf{g}\cdot{\mbox{\boldmath\scriptsize$\rho$}}}}{\sqrt{S}}
i\frac{\omega_{\mu}}{c} \,\hat{\epsilon}_{\textbf{g}} \left\{
\begin{array}{lc}
  A_{3\mu}e^{-\chi_{3\mu}(z-d/2)},          &  z > \frac{d}{2}\\
  A_{2\mu}e^{iq_{\mu}z}+B_{2\mu}e^{-iq_{\mu}z}, & |z|< \frac{d}{2} \\
  B_{1\mu}e^{ \chi_{1\mu}(z+d/2)},          &  z <-\frac{d}{2}\\
\end{array}
\right.
\end{equation}
\begin{equation}
\textbf{H}_{\mu}^{\mathrm{guided}}(\mbox{\boldmath$\rho$},z)=\frac
{e^{i\textbf{g}\cdot{\mbox{\boldmath\scriptsize$\rho$}}}} {\sqrt{S}}
\left\{
\begin{array}{lc}
  A_{3\mu}(\chi_{3\mu}\hat{g}+ig\hat{z})e^{-\chi_{3\mu}(z-d/2)}, & z>\frac{d}{2}\\
  A_{2\mu}i(-q_{\mu}\hat{g}+g\hat{z})e^{iq_{\mu}z}+B_{2g}i(q_{\mu}\hat{g}+g\hat{z})
    e^{-iq_{\mu}z}, & |z|<\frac{d}{2} \\
  B_{1\mu}(-\chi_{1\mu}\hat{g}+ig\hat{z})e^{ \chi_{1\mu}(z+d/2)}, & z<-\frac{d}{2}\\
\end{array}
\right.
\end{equation}
\end{widetext}
\end{small}
In the above expressions, $S$ is a normalization surface, which
disappears from the final results, while the magnetic field is
obtained from the electric field thorugh the Maxwell equation
${\bf H}(\textbf{r})=-\frac{ic}{\omega}\mbox{\boldmath$\nabla$}
\times {\bf E}(\textbf{r})\,$. Application of standard
transfer-matrix theory leads to the following relations between
coefficients (with the suffix $\mu$ being understood for
simplicity):
\begin{small}
\begin{eqnarray}
  A_2 & = & \frac{B_1}{2q}(q-i\chi_{1})e^{iqd/2} \label{eq:tecoeff1}\\
  B_2 & = & \frac{B_1}{2q}(q+i\chi_{1})e^{-iqd/2} \label{eq:tecoeff2}\\
  A_3 & = & \frac{B_1}{2q\chi_3}[q(\chi_{3}-\chi_{1})
  \cos(qd)+(q^2+\chi_{1}\chi_{3})\sin(qd)]\nonumber\\
  \label{eq:tecoeff3}
\end{eqnarray}
\end{small}
and to the well known implicit equation (\ref{eq:tepol}) of the
main text. The normalization integral (\ref{eq:orthonorm}) implies
the relation
\begin{small}
\begin{eqnarray}
&&\frac{\chi_1^2+g^2}{2\chi_1}|B_1|^2+\frac{\chi_3^2+g^2}{2\chi_3}|A_3|^2+
\nonumber\\
&& d\left[(g^2+q^2)(|A_2|^2+|B_2|^2)+ \right. \nonumber\\
&&\left. (g^2-q^2)(A_2^*B_2+B_2^*A_2)\frac{\sin(qd)}{qd}\right]=1
\end{eqnarray}
\end{small}
which, together with
Eqs.~(\ref{eq:tecoeff1})--(\ref{eq:tecoeff3}), determines all
coefficients $A_j$ and $B_j$.

The guided modes for TM polarization are characterized by the
following field profiles:
\begin{small}
\begin{widetext}
\begin{equation}
\textbf{H}_{\mu}^{\mathrm{guided}}(\mbox{\boldmath$\rho$},z)=\frac
{e^{i\textbf{g}\cdot{\mbox{\boldmath\scriptsize$\rho$}}}} {\sqrt{S}}
\hat{\epsilon}_{\textbf{g}} \left\{ \begin{array}{lc}
  C_{3\mu}e^{-\chi_{3\mu}(z-d/2)},          &  z > \frac{d}{2}\\
  C_{2\mu}e^{iq_{\mu}z}+D_{2\mu}e^{-iq_{\mu}z}, & |z|< \frac{d}{2} \\
  D_{1\mu}e^{ \chi_{1\mu}(z+d/2)},          &  z <-\frac{d}{2}\\
\end{array}
\right.
\end{equation}
\begin{equation}
\textbf{E}_{\mu}^{\mathrm{guided}}(\mbox{\boldmath$\rho$},z)=\frac
{e^{i\textbf{g}\cdot{\mbox{\boldmath\scriptsize$\rho$}}}}
{\sqrt{S}} i\frac{c}{\omega_{g}}\left\{
\begin{array}{lc}
  \frac{1}{\bar{\epsilon}_3}C_{3\mu}(\chi_{3\mu}\hat{g}+ig\hat{z})
  e^{-\chi_{3\mu}(z-d/2)}, & z>\frac{d}{2}\\
  \frac{1}{\bar{\epsilon}_2}[C_{2\mu}i(-q_{\mu}\hat{g}+g\hat{z})
  e^{ iq_{\mu}z}
  +D_{2\mu}i( q_{\mu}\hat{g}+g\hat{z})e^{-iq_{\mu}z}], & |z|<\frac{d}{2} \\
  \frac{1}{\bar{\epsilon}_1}D_{1\mu}(-\chi_{1\mu}\hat{g}+ig\hat{z})
  e^{ \chi_{1\mu}(z+d/2)}, & z<-\frac{d}{2}\\
\end{array}
\right.
\end{equation}
\end{widetext}
\end{small}
where the electric field follows from the magnetic field profile
through Eq.~(\ref{eq:efield}). The appropriate relations between
coefficients can be derived from
(\ref{eq:tecoeff1})-(\ref{eq:tecoeff3}) by the replacements
$A\rightarrow C$, $B\rightarrow D$ and
$\chi_1\rightarrow\chi_1/\bar{\epsilon}_1$, $q\rightarrow q
/\bar{\epsilon}_2$ (except in the trigonometric functions),
$\chi_3\rightarrow\chi_3/\bar{\epsilon}_3$. The normalization
integral of the magnetic field gives the condition
\begin{small}
\begin{equation}
\frac{|D_1|^2}{2\chi_1}+\frac{|C_3|^2}{2\chi_3} +
d\left[|C_2|^2+|D_2|^2+
(C_2^*D_2+D_2^*C_2)\frac{\sin(qd)}{qd}\right]=1 \, .
\end{equation}
\end{small}


\section{Matrix elements for photonic dispersion}
\label{app:guided-elements}

We calculate the matrix elements ${\cal{H}}_{\mu\nu}$ of Eq.
(\ref{eq:matrix-elements}) between guided modes of the effective
waveguide from the field profiles given in Appendix
\ref{app:guided-modes} for TE and TM polarizations. We adopt the
following notations:
$\mu\equiv(\textbf{k}+\textbf{G},\alpha)\equiv(\textbf{g},\alpha)$
and
$\nu\equiv(\textbf{k}+\textbf{G}',\alpha')\equiv(\textbf{g}',\alpha')$,
where $\textbf{k}$ is the Bloch-Floquet vector, $\textbf{G}$ and
$\textbf{G}'$ are reciprocal lattice vectors and $\alpha$,
$\alpha'$ are the indices of guided modes of the effective
waveguide at wavevectors $\textbf{k}+\textbf{G}$,
$\textbf{k}+\textbf{G}'$ respectively. There are four kinds of
matrix elements: TE-TE, TM-TM, TE-TM and TM-TE. The $z$ integral
in Eq. (\ref{eq:matrix-elements}) can be broken into three terms
over the regions 1,2,3 where the dielectric constant does not
depend on $z$, and can be expressed in terms of the following
integrals:
\begin{small}
\begin{eqnarray*}
I_3&\equiv&\int_{d/2}^{\infty}e^{-(\chi_{3\mu}+\chi_{3\nu})
(z-d/2)}\,\mathrm{d}z=(\chi_{3\mu}+\chi_{3\nu})^{-1},
\\
I_{2\pm}&\equiv&\int_{-d/2}^{d/2}e^{i(q_{\mu}\pm
q_{\nu})z}\,\mathrm{d}z=
\frac{\sin((q_{\mu}\pm q_{\nu})d/2)}{(q_{\mu}\pm q_{\nu})/2}, \\
I_1&\equiv&\int_{-\infty}^{-d/2}e^{(\chi_{1\mu}+\chi_{1\nu})(z+d/2)}
\,\mathrm{d}z=(\chi_{1\mu}+\chi_{1\nu})^{-1}.
\\
\end{eqnarray*}
\end{small}
The $xy$ integrals in Eq. (\ref{eq:matrix-elements}) yield the
inverse dielectric matrices in the three layers,
$\eta_j(\textbf{G},\textbf{G}')$, defined in Eq. (\ref{eq:eta}).
The matrix elements are then obtained as follows:
\begin{small}
\begin{eqnarray}
&& {\cal{H}}_{\mu\nu}^{\mathrm{TE-TE}}=
\left(\frac{\omega_{\mu}}{c}\right)^2
\left(\frac{\omega_{\nu}}{c}\right)^2
\hat{\epsilon}_{\textbf{g}}\cdot\hat{\epsilon}_{\textbf{g}'}
\nonumber \\
&& \times
\left\{(\bar{\epsilon}_1)^2{\eta}_1(\textbf{G},\textbf{G}')
B_{1\mu}^*B_{1\nu}I_1 \right. \nonumber \\
&& +(\bar{\epsilon}_3)^2{\eta}_3(\textbf{G},\textbf{G}')
A_{3\mu}^*A_{3\nu}I_3
+(\bar{\epsilon}_2)^2{\eta}_2(\textbf{G},\textbf{G}')\nonumber \\
&& \times \left.
\left[(A_{2\mu}^*A_{2\nu}+B_{2\mu}^*B_{2\nu})I_{2-}
    +  (A_{2\mu}^*B_{2\nu}+B_{2\mu}^*A_{2\nu})I_{2+} \right]
\right\}\,,\nonumber\\
\end{eqnarray}
\begin{eqnarray}
&& {\cal{H}}_{\mu\nu}^{\mathrm{TM-TM}}=
{\eta}_1(\textbf{G},\textbf{G}') D_{1\mu}^*D_{1\nu}
(\chi_{1\mu}\chi_{1\nu}\hat{g}\cdot\hat{g}'+gg')I_{1} \nonumber \\
&& +{\eta}_3(\textbf{G},\textbf{G}') C_{3\mu}^*C_{3\nu}
(\chi_{3\mu}\chi_{3\nu}\hat{g}\cdot\hat{g}'+gg')I_3 \nonumber
\\
&& +{\eta}_2(\textbf{G},\textbf{G}') \left[
(C_{2\mu}^*C_{2\nu}+D_{2\mu}^*D_{2\nu})\right. \nonumber \\
&&\times(q_{\mu}q_{\nu}\hat{g}\cdot\hat{g}'+gg') I_{2-}
+(C_{2\mu}^*D_{2\nu}+D_{2\mu}^*C_{2\nu})\nonumber \\
&&\left. \times(-q_{\mu}q_{\nu}\hat{g}\cdot\hat{g}'+gg') I_{2+}
 \right],
\end{eqnarray}
\begin{eqnarray}
&& {\cal{H}}_{\mu\nu}^{\mathrm{TE-TM}}=
\left(\frac{\omega_{\mu}}{c}\right)^2
\hat{\epsilon}_{\textbf{g}}\cdot\hat{g}' \left\{
-\bar{\epsilon}_1{\eta}_1(\textbf{G},\textbf{G}')
B_{1\mu}^*D_{1\nu}\chi_{1\nu}I_1 \right. \nonumber\\
&& +\bar{\epsilon}_3{\eta}_3(\textbf{G},\textbf{G}')
A_{3\mu}^*C_{3\nu}\chi_{3\nu}I_3
+i\bar{\epsilon}_2{\eta}_2(\textbf{G},\textbf{G}')q_{\nu}
\nonumber\\
&&\times\left[
(-A_{2\mu}^*C_{2\nu}+B_{2\mu}^*D_{2\nu})I_{2-}\right. \nonumber\\
&& \left.\left. +
(A_{2\mu}^*D_{2\nu}-B_{2\mu}^*C_{2\nu})I_{2+}\right] \right\},
\end{eqnarray}
\begin{eqnarray}
&& {\cal{H}}_{\mu\nu}^{\mathrm{TM-TE}}=
\left(\frac{\omega_{\nu}}{c}\right)^2
\hat{g}\cdot\hat{\epsilon}_{\textbf{g}'} \left\{
-\bar{\epsilon}_1{\eta}_1(\textbf{G},\textbf{G}')
D_{1\mu}^*B_{1\nu}\chi_{1\mu}I_1 \right. \nonumber\\
&& +\bar{\epsilon}_3{\eta}_3(\textbf{G},\textbf{G}')
C_{3\mu}^*A_{3\nu}\chi_{3\mu}I_3 \nonumber
\\ &&
-i\bar{\epsilon}_2{\eta}_2(\textbf{G},\textbf{G}')q_{\mu}
 \left[
(-C_{2\mu}^*A_{2\nu}+D_{2\mu}^*B_{2\nu})I_{2-}\right.\nonumber\\
&& + \left.\left. (D_{2\mu}^*A_{2\nu}-C_{2\mu}^*B_{2\nu})I_{2+}
\right]\right\}.
\end{eqnarray}
\end{small}

\section{Symmetry properties and TE/TM mixing}
\label{app:symmetry}

 If the PhC slab is invariant under reflection
through a mirror plane bisecting the slab (which we take as $z=0$),
the effective waveguide has the same reflection symmetry and its
eigenmodes can be classified as even or odd with respect to the
reflection operator $\hat{\sigma}_{xy}$ defined in the main text.
The spatial symmetries of the nonvanishing field components are
given in Table \ref{tab:fields}, assuming a wavevector along the
$\hat{x}$ direction. The guided mode frequencies are found from the
following implicit equations:
\begin{small}
\begin{eqnarray}
q\sin\frac{qd}{2}-\chi_1\cos\frac{qd}{2}=0 \;\;\; &\mathrm{TE},\,\sigma_{xy}=+1& \label{eq:te-even} \\
q\cos\frac{qd}{2}+\chi_1\sin\frac{qd}{2}=0 \;\;\; &\mathrm{TE},\,\sigma_{xy}=-1& \label{eq:te-odd} \\
\frac{q}{\bar{\epsilon}_2}\cos\frac{qd}{2}+\frac{\chi_1}{\bar{\epsilon}_1}\sin\frac{qd}{2}=0
\;\;\; &\mathrm{TM},\,\sigma_{xy}=+1& \label{eq:tm-even} \\
\frac{q}{\bar{\epsilon}_2}\sin\frac{qd}{2}-\frac{\chi_1}{\bar{\epsilon}_1}\cos\frac{qd}{2}=0
\;\;\; &\mathrm{TM},\,\sigma_{xy}=-1& \label{eq:tm-odd}
\end{eqnarray}
\end{small}
Notice that the first-order TE and TM modes of the effective
waveguide are described by Eqs.~(\ref{eq:te-even}) and
(\ref{eq:tm-odd}), respectively. By keeping only $\sigma_{xy}=+1$
($\sigma_{xy}=-1$) basis states in the guided-mode expansion,
photonic eigenmodes which are even (odd) with respect to
horizontal mirror symmetry are obtained, and the computational
effort in solving the eigenvalue equation (\ref{eq:linear}) is
reduced.
\begin{table}[t]
\caption{Spatial symmetry of the electric and magnetic field
components of an eigenmode of a symmetric waveguide with respect
to the mirror symmetry operation $\hat{\sigma}_{xy}$. The
wavevector is assumed to lie along the $\hat{x}$ direction.
Vanishing components are denoted by ``*''.}
\begin{tabular}{c|cccccc}
                               & $E_x$ & $E_y$ & $E_z$ & $H_x$ & $H_y$ & $H_z$
                               \\ \hline
$\mathrm{TE},\,\sigma_{xy}=+1$ &   *  &   $+$ &    *   &  $-$  &   *   &  $+$ \\
$\mathrm{TE},\,\sigma_{xy}=-1$ &   *  &   $-$ &    *   &  $+$  &   *   &  $-$ \\
$\mathrm{TM},\,\sigma_{xy}=+1$ &  $+$ &    *  &   $-$  &   *   &  $-$  &   *  \\
$\mathrm{TM},\,\sigma_{xy}=-1$ &  $-$ &    *  &   $+$  &   *   &  $+$  &   *  \\
\hline
\end{tabular}
\label{tab:fields} \vspace*{1cm}
\end{table}

In general, a photonic eigenmode in a PhC slab is a linear
combination of TE- and TM-polarized basis states and all six
components of the electric and magnetic field are nonvanishing.
Nevertheless, low-lying photonic modes are often dominated by the
lowest-order guided mode of the effective waveguide. In this
situation, $\sigma_{xy}=+1$ states are dominated by the TE guided
modes described by Eq.~(\ref{eq:te-even}) and can be called
``quasi-TE'', while $\sigma_{xy}=-1$ states are dominated by the TM
guided modes described by Eq.~(\ref{eq:tm-odd}) and can be called
``quasi-TM''. This widespread terminology is useful and appropriate
for large wavevectors, when the dominant field components are the
spatially even ones: making reference to Table \ref{tab:fields}, and
since the dielectric modulation $\epsilon_j(xy)$ couples guided
modes with all wavevector directions in the 2D plane, the dominant
components are $E_x,E_y,H_z$ (for $\sigma_{xy}=+1$ or TE-like modes)
or $H_x,H_y,E_z$ (for $\sigma_{xy}=-1$ or TM-like modes). However,
the terminology becomes inadequate when the PhC slab is multimode,
as it always happens at sufficiently high frequencies. If the
waveguide is not symmetric, any PhC slab mode is a linear
combination of basis states arising from all equations
(\ref{eq:te-even})--(\ref{eq:tm-odd}) and even the lowest-order
TE/TM guided modes of the effective waveguide are mixed.

\section{Radiation modes of the effective waveguide}
\label{app:rad-modes}

In this Appendix we give the radiation modes of the effective
waveguide which are outgoing in either the lower or the upper
cladding (see Fig.~\ref{fig:scatt-states}). Each of these radiation
modes is labelled by the wavevector \textbf{g} in the 2D plane, the
frequency $\omega$, and the polarization. In the following the
quantum numbers $\textbf{g}, \omega$ will be understood in order to
simplify the notations. Instead of distinguishing between real and
imaginary components of the wavevectors in the three regions as in
Eq.~(\ref{eq:chi-and-q}), we define three complex wavevectors as
\begin{small}
\begin{equation}
q_{j}=\left(\bar{\epsilon}_j\frac{\omega^2}{c^2}-g^2\right)^{1/2},\;j=1,2,3.
\label{eq:q}
\end{equation}
\end{small}
While the wavevector $q_{2}$ in the core region is always real,
the wavevectors $q_{1},q_{3}$ in the cladding regions can be
either real or purely imaginary according to the values of the
dielectric constants and of $\textbf{g}, \omega$. If the effective
waveguide is symmetric ($\bar{\epsilon}_1=\bar{\epsilon}_3$) all
radiation modes have real $q_1,q_3$, while if
$\bar{\epsilon}_1>\bar{\epsilon}_3$ there is a frequency region
$cg/\sqrt{\bar{\epsilon_1}}<\omega<cg/\sqrt{\bar{\epsilon}_3}$ in
which $q_1$ is real but $q_3$ is imaginary: i.e., a quasi-guided
mode in this region of $k$-$\omega$ space is diffracted to the
lower (but not to the upper) cladding. This phenomenon is
automatically taken into account by the Kronecker
$\theta$-function in the DOS formula (\ref{eq:dos}), as the
photonic DOS vanishes when $q_{j}$ is imaginary. Therefore we
shall write down the electric and magnetic field components of
radiative waveguide modes assuming that all $q_j$'s are real - if
one of them is purely imaginary, that radiation mode does not
carry an energy flux and it does not contribute to scattering
loss.

With this proviso, the radiation modes of the effective waveguide
for TE polarization in the three regions $j=1,2,3$ can  be written
as follows (setting $z_1=-d/2$, $z_2=0$, $z_3=d/2$ and with the
time dependence $e^{-i\omega_{g}t}$ being understood):
\begin{small}
\begin{equation}
\textbf{E}_{j}^{\mathrm{rad}}(\mbox{\boldmath$\rho$},z)=\frac
{e^{i\textbf{g}\cdot{\mbox{\boldmath\scriptsize$\rho$}}}} {\sqrt{S}}
i\hat{\epsilon}_{\textbf{g}} \left[ W_j e^{iq_{j}(z-z_j)} + X_j
e^{-iq_{j}(z-z_j)} \right],
\end{equation}
\begin{eqnarray}
\textbf{H}_{j}^{\mathrm{rad}}(\mbox{\boldmath$\rho$},z)&=&\frac
{e^{i\textbf{g}\cdot{\mbox{\boldmath\scriptsize$\rho$}}}}
{\sqrt{S}} i\frac{c}{\omega} \left[ (g\hat{z}-q_j\hat{g}) W_j
e^{iq_{j}(z-z_j)}\right. \nonumber\\
&+& \left. (g\hat{z}+q_j\hat{g})X_j e^{-iq_{j}(z-z_j)} \right]\,.
\end{eqnarray}
\end{small}
For the scattering state which is outgoing in the lower cladding
(Fig.~\ref{fig:scatt-states}a), $W_1=0$ and the normalization
condition (\ref{eq:orthonorm}) determines the coefficient of the
outgoing component\cite{note:scattering} as
$X_1=1/\sqrt{\bar{\epsilon_1}}$.  All other coefficients are then
found from a standard transfer-matrix calculation, which yields
\begin{small}
\begin{equation}
\left[ \begin{array}{c} W_2 \\ X_2 \\ \end{array} \right]=
\frac{1}{2q_2}\left[
\begin{array}{cc} (q_2+q_1)e^{iq_2d/2} & (q_2-q_1)e^{iq_2d/2}\\
(q_2-q_1)e^{-iq_2d/2} & (q_2+q_1)e^{-iq_2d/2}\\ \end{array}
\right] \left[ \begin{array}{cc} W_1 \\ X_1 \\ \end{array} \right]
\label{eq:transfer-matrix-1}
\end{equation}
\end{small}
and
\begin{small}
\begin{equation}
\left[ \begin{array}{c} W_3 \\ X_3 \\ \end{array} \right]=
\frac{1}{2q_3}\left[
\begin{array}{cc} (q_3+q_2)e^{iq_2d/2} & (q_3-q_2)e^{-iq_2d/2}\\
(q_3-q_2)e^{iq_2d/2} & (q_3+q_2)e^{-iq_2d/2}\\ \end{array} \right]
\left[ \begin{array}{cc} W_2 \\ X_2 \\ \end{array} \right].
\label{eq:transfer-matrix-2}
\end{equation}
\end{small}
On the other hand, for the state outgoing in the upper cladding
(Fig.~\ref{fig:scatt-states}b), $X_3=0$ and the proper normalization
condition\cite{note:scattering} is $W_3=1/\sqrt{\bar{\epsilon_3}}$.
The other coefficients can then be found from the inverse relations
of (\ref{eq:transfer-matrix-1}), (\ref{eq:transfer-matrix-2}).

For what concerns TM-polarized radiation modes, the field profiles
are given by
\begin{small}
\begin{equation}
\textbf{H}_{j}^{\mathrm{rad}}(\mbox{\boldmath$\rho$},z)=\frac
{e^{i\textbf{g}\cdot{\mbox{\boldmath\scriptsize$\rho$}}}} {\sqrt{S}}
\hat{\epsilon}_{\textbf{g}} \left[ Y_j e^{iq_{j}(z-z_j)} + Z_j
e^{-iq_{j}(z-z_j)} \right],
\end{equation}
\begin{eqnarray}
\textbf{E}_{j}^{\mathrm{rad}}(\mbox{\boldmath$\rho$},z)&=&-\frac
{e^{i\textbf{g}\cdot{\mbox{\boldmath\scriptsize$\rho$}}}}
{\sqrt{S}} \frac{c}{\bar{\epsilon}_j\omega} \left[
(g\hat{z}-q_j\hat{g}) Y_j e^{iq_{j}(z-z_j)}\right. \nonumber\\
&+& \left. (g\hat{z}+q_j\hat{g})Z_j e^{-iq_{j}(z-z_j)} \right].
\end{eqnarray}
\end{small}
For the scattering state outgoing in the lower cladding (Fig.
\ref{fig:scatt-states}a), $Y_1=0$ and the normalization condition
(\ref{eq:orthonorm}) determines the coefficient of the outgoing
component as $Z_1=1$. For the state outgoing in the upper cladding
(Fig.~\ref{fig:scatt-states}b), $Z_3=0$ and the proper normalization
condition is $Y_3=1$. All other coefficients can be found from
transfer-matrix theory, the relevant expressions being obtained from
(\ref{eq:transfer-matrix-1}), (\ref{eq:transfer-matrix-2}) by the
replacements $W\rightarrow Y$, $X\rightarrow Z$ and $q_j\rightarrow
q_j/\bar{\epsilon}_j$ (except in the exponential functions).

\section{Matrix elements for diffraction losses}
\label{app:rad-elements}

We calculate the matrix elements (\ref{eq:guided-rad-gme}) between
guided and radiation modes of the effective waveguide, to be used in
formula (\ref{eq:k-rad-gme}) for the loss calculation. The
expressions are similar to those of Appendix
\ref{app:guided-elements}, but care must be taken to distinguish
between the different quantum numbers. Guided modes are labelled by
the wavevector $\textbf{k}+\textbf{G}\equiv\textbf{g}$ and by the
index $\alpha$, which are combined as before into a single index
$\mu$. Radiation modes are labelled by the wavevector
$\textbf{k}+\textbf{G}'\equiv\textbf{g}'$, the frequency $\omega$,
the polarization $\lambda$, and an additional index which specifies
whether the mode is outgoing in medium 1 or 3: all these quantum
numbers will be grouped into a single index $r$. Like for the matrix
elements between guided modes, there are four possible combinations
of polarizations. The $z$ integral can be broken into three terms
over the regions 1,2,3 and it can be expressed in terms of the
following integrals:
\begin{small}
\begin{eqnarray*}
I_{3\pm}&\equiv&\int_{d/2}^{\infty}e^{-(\chi_{3\mu}\pm
iq_{3r})(z-d/2)}\,\mathrm{d}z=(\chi_{3\mu}\pm iq_{3r})^{-1},
\\
I_{2\pm}&\equiv&\int_{-d/2}^{d/2}e^{i(q_{\mu}\pm
q_{2r})z}\,\mathrm{d}z=
\frac{\sin((q_{\mu}\pm q_{2r})d/2)}{(q_{\mu}\pm q_{2r})/2}, \\
I_{1\pm}&\equiv&\int_{-\infty}^{-d/2}e^{(\chi_{1\mu}\pm
iq_{1r})(z+d/2)}\,\mathrm{d}z=(\chi_{1\mu}\pm iq_{1r})^{-1}.
\end{eqnarray*}
\end{small}
The matrix elements between guided and radiative modes are found
as follows:
\begin{small}
\begin{eqnarray}
&& {\cal{H}}_{\mu,r}^{\mathrm{TE-TE}}=
\left(\frac{\omega_{\mu}}{c}\right)^2 \frac{\omega_{r}}{c} \,
\hat{\epsilon}_{\textbf{g}}\cdot\hat{\epsilon}_{\textbf{g}'}
\left\{(\bar{\epsilon}_1)^2{\eta}_1(\textbf{G},\textbf{G}')
B_{1\mu}^* \right.\nonumber\\
&& \times(W_{1r}I_{1+}+X_{1r}I_{1-})
+(\bar{\epsilon}_3)^2{\eta}_3(\textbf{G},\textbf{G}')
A_{3\mu}^* \nonumber\\
&& \times (W_{3r}I_{3-}+X_{3r}I_{3+})
+(\bar{\epsilon}_2)^2{\eta}_2(\textbf{G},\textbf{G}')\nonumber\\
&& \times\left.\left[(A_{2\mu}^*W_{2r}+B_{2\mu}^*X_{2r})I_{2-}
    + (A_{2\mu}^*X_{2r}+B_{2\mu}^*W_{2r})I_{2+} \right]
\right\}\, \nonumber\\
\end{eqnarray}
\begin{eqnarray}
&&{\cal{H}}_{\mu,r}^{\mathrm{TM-TM}}=
{\eta}_1(\textbf{G},\textbf{G}') D_{1\mu}^*
[(gg'+i\chi_{1\mu}q_{1r}\hat{g}\cdot\hat{g}')Y_{1r}I_{1+}\nonumber\\
&& +(gg'-i\chi_{1\mu}q_{1r}\hat{g}\cdot\hat{g}')Z_{1r} I_{1-}]\nonumber\\
&& +{\eta}_3(\textbf{G},\textbf{G}') C_{3\mu}^*
[(gg'-i\chi_{3\mu}q_{3r}\hat{g}\cdot\hat{g}')Y_{3r}I_{3-}\nonumber\\
&& +(gg'+i\chi_{3\mu}q_{3r}\hat{g}\cdot\hat{g}')Z_{3r}I_{3+}]\nonumber\\
&& +{\eta}_2(\textbf{G},\textbf{G}')
[(C_{2\mu}^*Y_{2r}+D_{2\mu}^*Z_{2r})\nonumber\\
&& \times(gg'+q_{\mu}q_{2r}\hat{g}\cdot\hat{g}') I_{2-}\nonumber\\
&& +(C_{2\mu}^*Z_{2r}+D_{2\mu}^*Y_{2r})
(gg'-q_{\mu}q_{2r}\hat{g}\cdot\hat{g}')I_{2+}]\, ,
\end{eqnarray}
\begin{eqnarray}
&&{\cal{H}}_{\mu,r}^{\mathrm{TE-TM}}=
i\left(\frac{\omega_{\mu}}{c}\right)^2
\hat{\epsilon}_{\textbf{g}}\cdot\hat{g}' \left\{
\bar{\epsilon}_1{\eta}_1(\textbf{G},\textbf{G}')q_{1r}
B_{1\mu}^*\right. \nonumber\\
&& \times(-Y_{1r}I_{1+}+Z_{1r}I_{1-}) \nonumber\\
&& +\bar{\epsilon}_3{\eta}_3(\textbf{G},\textbf{G}')q_{3r}
A_{3\mu}^*(-Y_{3r}I_{3-}+Z_{3r}I_{1+})\nonumber\\
&& +\bar{\epsilon}_2{\eta}_2(\textbf{G},\textbf{G}')q_{2r} \left[
(-A_{2\mu}^*Y_{2r}+B_{2\mu}^*Z_{2r})I_{2-}\right.\nonumber\\
&& \left.\left. +
(A_{2\mu}^*Z_{2\nu}-B_{2\mu}^*Y_{2r})I_{2+}\right] \right\} \,,
\end{eqnarray}
\begin{eqnarray}
&& {\cal{H}}_{\mu,r}^{\mathrm{TM-TE}}= \frac{\omega_{r}}{c}
\hat{g}\cdot\hat{\epsilon}_{\textbf{g}'}
\left\{-\bar{\epsilon}_1{\eta}_1(\textbf{G},\textbf{G}')
\chi_{1\mu}D_{1\mu}^*\right. \nonumber\\
&& \times(W_{1r}I_{1+}+X_{1r}I_{1-}) \nonumber\\ &&
+\bar{\epsilon}_3{\eta}_3(\textbf{G},\textbf{G}')
\chi_{3\mu}C_{3\mu}^*(W_{3r}I_{3-}+X_{3r}I_{3+}) \nonumber\\
&& -i\bar{\epsilon}_2{\eta}_2(\textbf{G},\textbf{G}')q_{\mu}
\left[(D_{2\mu}^*X_{2r}-C_{2\mu}^*W_{2r})I_{2-} \right. \nonumber\\
&& \left. \left. +(D_{2\mu}^*W_{2r}-C_{2\mu}^*X_{2r})I_{2+}
\right] \right\} \, .
\end{eqnarray}
\end{small}

\section{Fourier transform of triangular holes}
\label{app:four-trans}

We give here the complex Fourier transform of the dielectric
constant in the unit cell for the triangular lattice of triangular
holes, to be used for the calculations shown in
Secs.~\ref{sec:trianglatt} and \ref{sec:defects}. Let us refer to
the geometry shown in Fig.~\ref{fig:scheme-tria}, the side of the
triangle and the lattice constant being denoted by $L$ and $a$,
respectively. The triangle is made of a medium of dielectric
constant $\epsilon_1$ (in the case of an air hole, $\epsilon_1=1$)
embedded in a background with dielectric constant $\epsilon_2$.
The origin of coordinates in the unit cell is taken to be at the
center of the triangle. The filling factor of the triangle is
\begin{small}
\begin{equation}
f=\frac{\sqrt{3}L^2}{4A},
\end{equation}
\end{small}
where $A=\sqrt{3}a^2/2$ is the unit-cell area. The Fourier
transform is written as
\begin{small}
\begin{equation}
\epsilon(\textbf{G})=\frac{1}{A}\int_{\mathrm{cell}}
\epsilon(\mbox{\boldmath$\rho$})e^{-i\textbf{G}
\cdot\mbox{\boldmath\small$\rho$}}\,\mathrm{d}\mbox{\boldmath$\rho$}.
\end{equation}
\end{small}
For zero reciprocal lattice vector it is obviously given by:
\begin{small}
\begin{equation}
\epsilon(\textbf{G}=0)= f\epsilon_1+(1-f)\epsilon_2.
\end{equation}
\end{small}
For nonzero reciprocal lattice vector it is calculated as
\begin{small}
\begin{equation}
\epsilon(\textbf{G}\neq0)=f(\epsilon_1-\epsilon_2)[I(G_x,G_y)+I(-G_x,G_y)],
\end{equation}
\end{small}
where
\begin{small}
\begin{equation}
I(G_x,G_y=0)=\frac{1}{g_x^2}(1-e^{-ig_x})-\frac{i}{g_x},
\end{equation}
\begin{equation}
I(G_x,G_y\neq0)=\frac{i}{g_y}e^{i(\frac{g_y}{3}-\frac{g_x}{2})}\left[
e^{-i\frac{g_y}{2}}j_0\left(\frac{g_x-g_y}{2}\right)-j_0\left(\frac{g_x}{2}\right)\right],
\end{equation}
\end{small}
with $g_x=G_xL/2$, $g_y=\sqrt{3}G_yL/2$. In this work, the
triangles are oriented in such a way that one of the triangle'
sides is along the $x$-axis as in Fig.~\ref{fig:scheme-tria}
(i.e., along the $\Gamma$-K direction of the triangular lattice):
if other orientations have to be considered, the simplest way is
to apply a 2D rotation matrix to the vector \textbf{G} before
calculating the Fourier transform with the above formulas.


\begin{thebibliography}{99}
\bibitem{note:semi-infinite}
In order to derive the formulas of this Section, it is in fact
necessary to consider a finite thickness ${\cal{L}}$ of the
cladding layers and to take the limit ${\cal{L}}\rightarrow
\infty$ of the resulting expressions. These details will be
omitted for simplicity.
\bibitem{note:alpha}
Notice that the index $\alpha$ takes into account both the order
and the polarization of the guided mode. For example, if two TE
and two TM guided modes are kept in the expansion, there are four
distinct values of $\alpha$.
\bibitem{note:outgoing}
It is shown in Ref.~\protect\onlinecite{ochiai01-pert} that only
the outgoing components of the radiation modes contribute to
Im($\omega$), as physically expected.
\bibitem{note:normalization}
This can be shown formally by introducing a normalization box of
width ${\cal{L}}$ in the $z$ direction. When ${\cal{L}}\gg d$, the
contributions of the cladding regions to the normalization
integral extend over a very large thickness and the core region
becomes irrelevant.  On the other hand, the contributions of core
and claddings to the matrix element
(\protect\ref{eq:guided-rad-gme}) are of the same order in
$\cal{L}$ (they are both proportional to $({\cal{L}})^{-1/2}$)
because the integrals in the cladding regions are exponentially
damped by the field of the guided modes. Since the fictitious
width ${\cal{L}}$ disappears from the final result for
$\mathrm{Im}(\omega)$, it has been set equal to unity in
Eq.~(\ref{eq:dos}) and in Appendix~\ref{app:rad-modes} for the
sake of simplicity.
\bibitem{note:scattering}
This result can be derived in an equivalent way from the electric
field profile and the normalization condition
(\protect\ref{eq:orthonorm-efield}) for the electric field.
\bibitem{note:SHG}
Notice that in materials with the zincblende structure, the
symmetry of the nonlinear susceptibility tensor allows for SHG
with a TE-polarized pump and a TM-polarized harmonic in a
(001)-oriented slab.
\bibitem{note:sakoda}
The results of Fig.~\protect\ref{fig:bulk-loss}a,c are similar to
those reported in Ref.~\protect\onlinecite{ochiai01-fdtd}, which
are calculated with the FDTD method for a structure with slightly
lower air fraction.
\bibitem{note:proploss}
Notice that propagation losses, $\alpha_{\mathrm{loss}}$, of
line-defect modes can be derived from the imaginary part of
frequency through the relation
$\alpha_{\mathrm{loss}}=2\mathrm{Im}(k)=2\mathrm{Im}(\omega)/v_{g}$,
where $v_{g}$ is the group velocity of the mode.
\end{thebibliography}
\end{document}